\documentstyle[preprint,aps,version2]{revtex}
\begin{document}
\font\fortssbx=cmssbx10 scaled \magstep2
\hbox to \hsize{
\setlength{\footskip}{1.5cm}
\frenchspacing
\everymath={\displaystyle}
  \def\thebibliography#1{{\bf{References}}\list
 {[\arabic{enumi}]}{\settowidth\labelwidth{[#1]}\leftmargin\labelwidth
   \advance\leftmargin\labelsep
   \usecounter{enumi}}
   \def\newblock{\hskip .11em plus .33em minus -.07em}
   \sloppy
   \sfcode`\.=1000\relax}
  \let\endthebibliography=\endlist
\hskip.5in \raise.1in\hbox{\fortssbx University of Wisconsin - Madison}
\hfill$\vcenter{\hbox{\bf MAD/PH/857}
                \hbox{\bf UCD--94--39}
                \hbox{December 1994}}$ }
\vspace{.5in}

\begin{title}
Angular Distributions of Drell-Yan Lepton Pairs at the Tevatron:\\
Order $\alpha_s^2$ Corrections and Monte Carlo Studies
\end{title}

\author{E.~Mirkes$^a$ and J.~Ohnemus$^b$}

\begin{instit}
$^a$Physics Department, University of Wisconsin, Madison, WI 53706, USA\\
$^b$Physics Department, University of California, Davis, CA 95616, USA
\end{instit}

\begin{abstract}
We investigate the angular distribution of the lepton pair  in the
process $p \bar{p} \rightarrow \gamma^{\ast} + X \rightarrow \ell^+
\ell^- + X$, where the virtual photon is produced at high transverse
momentum. The angular distribution of the leptons is very sensitive to
possible  nonperturbative effects, such as  a nontrivial vacuum
structure of QCD, and offers a good chance to test such effects. We
present complete ${\cal O}(\alpha_s^2)$  calculations of the decay
lepton distributions in the lepton pair rest frame.  An order ${\cal
O}(\alpha_s)$ Monte Carlo study of the  lepton angular distributions,
with acceptance cuts and energy resolution smearing applied to the
leptons, is also presented.
\end{abstract}
\thispagestyle{emtpy}
\newpage

\section{INTRODUCTION}

The production of a virtual photon at hadron colliders, along with the
subsequent decay  into a lepton pair, provides a unique opportunity
for testing  perturbative Quantum Chromodynamics (QCD). In particular,
the  measurement of the angular distribution of the decay leptons
provides a detailed test of the production mechanism of the virtual
photon. Experimental studies  of the angular distribution of the
lepton pair have been made by the NA10 collaboration at
CERN~\cite{na10} and the Chicago-Iowa-Princeton collaboration at
Fermilab~\cite{cip}.  The data from these experiments do not agree
with the predictions of the QCD improved parton model. An
investigation of the decay lepton distribution at the Tevatron center
of mass energy
would certainly be interesting  in view of these results. In this
paper, we present next-to-leading order QCD predictions for the
angular distributions of the lepton pair in the process
$p\bar{p}\rightarrow\gamma^*+X\rightarrow \ell^+\ell^-+X$ at Tevatron
energies together with a leading order Monte Carlo study in which
acceptance cuts and energy resolution smearing are applied to the
leptons.

A basic assumption in the calculation of the Drell-Yan
process~\cite{dy} is the factorization hypothesis, which asserts that the
hadronic cross section can be calculated by convoluting the cross
section for the corresponding parton level reaction, $q \bar q \to
\gamma^{\ast} \to \ell^+ \ell^-$,  with the parton distribution
functions of the initial state hadrons. Great theoretical efforts have
been made to prove this  factorization assumption within perturbative
QCD~\cite{factorization1}. However, in Ref.~\cite{nachtmann1} it has
been argued that nonperturbative QCD effects may spoil the
factorization. Nonperturbative vacuum fluctuations could generate a
background field (like the color domains  described in the first paper
of Ref.~\cite{nachtmann1}) which could  induce spin and color
correlations between the partons of the  incoming hadrons.  These
effects need {\it not} vanish in the high energy limit,  {\it i.e.},
they can be ``leading twist'' effects  (for reviews see
Ref.~\cite{rev}).

It has recently been shown~\cite{nachtmann2} that the decay lepton
distribution of a vector boson produced in hadronic  collisions is
very sensitive to nonperturbative effects and offers a good chance to
test the factorization hypothesis. As mentioned earlier, the
experimental results in Refs.~\cite{na10}  and~\cite{cip} for the
decay lepton distribution from a virtual photon produced in
pion--nucleon scattering  are not in agreement with next-to-leading
order QCD predictions~\cite{nachtmann2} and therefore give an
indication that factorization may be violated. In
Ref.~\cite{nachtmann2} it has been shown that these results can be
explained by transverse  spin correlations between the initial state
partons, which may be induced by the nontrivial vacuum structure in
QCD. Pion bound state effects have been discussed in
Ref.~\cite{brodsky} as  another possible explanation for the
discrepancy.

It would clearly be interesting to check whether the decay lepton
distributions from gauge bosons produced at the Tevatron are in
agreement with the factorization assumption. To do this test, reliable
theoretical predictions for the decay distributions are mandatory. In
this paper, we present complete next-to-leading order [${\cal
O}(\alpha_s^2)$] predictions for the angular distributions of the
leptons produced via  the decay of a high $p_T^{}$ virtual photon,
{\it i.e.}, the  Drell-Yan process  $p \bar p \to \gamma^* + X \to
\ell^+ \ell^- + X$ is calculated to ${\cal O}(\alpha_s^2)$.  The
calculation is based on the assumption of factorization, {\it i.e.},
the calculation is done in the standard QCD-improved parton model.
When the virtual photon is produced with no transverse momentum, the
zero order Drell-Yan subprocess,  $q \bar q \to \gamma^* \to \ell^+
\ell^-$, predicts a $1 + \cos^2 \theta$ distribution for the leptons,
where $\theta$ is the scattering angle in the parton center of mass
frame (the $\gamma^*$ rest frame). For virtual photons produced with
transverse momentum  (balanced by additional gluons or quarks), the
event plane spanned by the beam and virtual photon momentum directions
provides a convenient reference plane for studying the angular
distributions of the decay leptons. The angular distribution now has
the general form
\begin{equation}
{d\sigma \over d\phi \, d\cos\theta} \sim (1+\cos\theta^2)
  + \frac{1}{2} \, A_{0} \, (1-3\cos^{2}\theta)
  +  A_{1}  \, \sin 2\theta \cos\phi
  + \frac{1}{2} \, A_{2}  \, \sin^{2}\theta\cos 2\phi \>,
\end{equation}
where $\theta$ and $\phi$ denote the polar and azimuthal angles of the
decay leptons in the virtual photon rest frame. The coefficients $A_i$
are functions of the transverse momentum and rapidity of the virtual
photon (measured in the laboratory frame) and vanish in the limit
$p_T^{}(\gamma^*)\rightarrow 0$.

The remainder of this paper is organized as follows. In Sec.~II we
discuss the formalism for describing the angular distributions of the
decay leptons, give an overview of the ${\cal O}(\alpha_s^2)$
calculation of the process $p \bar p \to \gamma^* + X \to \ell^+
\ell^- + X$, and define two choices for the $z$-axis in the lepton
pair rest frame. In Sec.~III we present  ${\cal O}(\alpha_s^2)$
numerical  results for the coefficients $A_i$ and show  that the QCD
corrections to the coefficients are fairly small. This is because the
coefficients $A_i$  are  ratios of helicity cross sections  [see
Eq.~(\ref{aintr})] and the QCD corrections tend to   cancel in these
ratios.  A leading order [${\cal O}(\alpha_s)$]  Monte Carlo study of
the decay lepton distributions for the Tevatron center of mass energy,
with  typical acceptance cuts and energy resolution smearing  applied
to the leptons, is also given. Conclusions and summary are given in
Sec.~IV. Finally, there is an appendix containing next-to-leading
order  matrix elements which  contribute to virtual photon production
but have not appeared in the literature.

\section{FORMALISM}

The formalism used in our calculations is discussed in this section.
First the methodology for describing the angular distributions of the
leptons is reviewed, then an overview of the ${\cal O}(\alpha_s^2)$
calculation of the  process $p \bar{p} \rightarrow \gamma^{\ast} + X
\rightarrow \ell^{+} \ell^{-} + X$  is given, and finally, two choices
for the $z$-axis in the lepton pair rest frame are defined.

\subsection{Angular distributions}

We consider the angular distributions of the leptons coming from the
leptonic decay of a virtual photon  produced with non-zero transverse
momentum in high energy proton-antiproton collisions. For definiteness
we take
\begin{equation}
 p(P_{1}) + \bar{p}(P_{2}) \rightarrow \gamma^{\ast}(Q) + X \rightarrow
\ell^{+}(\ell_{1}) + \ell^{-}(\ell_{2}) + X \>,
\label{ppbarw}
\end{equation}
where the quantities in parentheses denote the four-momenta of the
particles. At leading order [${\cal O}(\alpha_s$)] the parton
subprocesses
\begin{equation}
q + \bar{q} \rightarrow \gamma^{\ast} + g  \>, \qquad \qquad
q + g       \rightarrow \gamma^{\ast} + q  \>,
\label{loprocess}
\end{equation}
contribute to high $p_{T}^{}$ virtual photon production.

In the parton model the hadronic cross section is obtained by folding
the hard scattering parton level cross section with the respective
parton densities:
\begin{equation}
\frac{d\sigma^{h_{1}h_{2}}}{ dQ^2\,dp_{T}^{2}\,dy\,d\Omega^{\ast} } =
\sum_{a,b}
\int\,dx_{1}\,dx_{2}\,
f_{a}^{h_{1}}(x_{1},\mu_F^{2})\,
f_{b}^{h_{2}}(x_{2},\mu_F^{2})
\,
\frac{{s}\,d{\hat{\sigma}}_{ab}}{ dQ^2\,d{t}\,d{u}\,d\Omega^{\ast}}\,
\left( x_{1}P_{1},x_{2}P_{2},\alpha_s(\mu_R^{2}) \right) ,
\label{wqhad}
\end{equation}
where the sum is over $a,b=q, \bar{q}, g$. $\,f_{a}^{h}(x,\mu^2_F)$ is
the probability density for finding parton $a$ with momentum fraction
$x$ in hadron $h$ when it is probed at scale $\mu_F^{2}$. The parton
level  cross section for the subprocesses in Eq.~(\ref{loprocess}) are
denoted by $d{\hat{\sigma}}_{ab}$.  Note that in the framework of the
parton model the incoming partons are assumed to be unpolarized in
spin and color (for unpolarized initial state hadrons). Furthermore,
one neglects the transverse momenta of the incoming partons. (This is
in contrast to the model discussed in Ref.~\cite{nachtmann2} where it
is assumed that a  nontrivial structure of the QCD vacuum correlates
the spins and momenta of the incoming partons. Given such a
correlation, it has been shown that these correlations can drastically
affect the polarization of the produced boson~\cite{nachtmann2}.)

Denoting hadron level and parton level quantities by upper and lower
case characters, respectively,  the hadron and parton level Mandelstam
variables are  defined by
\begin{equation}
S = (P_1 + P_2)^2 \>, \qquad T = (P_1 - Q)^2 \>, \qquad U = (P_2 - Q)^2 \>,
\end{equation}
and
\begin{equation}
  \begin{array}{lclcl}
\label{kleinmandeldef}
s&=&(p_{1}+p_{2})^{2}&=&x_{1}x_{2}S \>, \\[2mm]
t&=&(p_{1}-Q)^{2}    &=&x_{1}(T-Q^2)+Q^2 \>, \\[2mm]
u&=&(p_{2}-Q)^{2}    &=&x_{2}(U-Q^2)+Q^2 \>,
  \end{array}
\end{equation}
where $p_1 = x_1 P_1$ and $p_2 = x_2 P_2$. The rapidity $y$ of the
virtual photon in the laboratory frame can be written
\begin{equation}
y=\frac{1}{2}\ln\left(\frac{Q^2-U}{Q^2-T}\right) \>,
\end{equation}
and the transverse momentum $p_T^{}$ of the virtual photon is related
to the Mandelstam variables via
\begin{equation}
p_{T}^{2}=\frac{(Q^2-U)(Q^2-T)}{S}-Q^2 \>.
\end{equation}
The angles $\theta$ and  $\phi$ in $d\Omega^{\ast} =
d\cos\theta\,d\phi$ are the polar and azimuthal decay angles of the
leptons in the virtual photon rest frame, measured  with respect to a
coordinate system to be described later. The angular dependance of the
differential cross section in Eq.~(\ref{wqhad}) can be written (see
Ref.~\cite{npb} for details)
\begin{eqnarray}
\frac{16\pi}{3}\frac{d{\sigma}}{dQ^2\,dp_T^2\,dy\,d\cos\theta\,d\phi} =
&\phantom{+}& \frac{d{\sigma}^{U+L}}{dQ^2\,dp_T^2\,dy}\,(1+\cos^2\theta)
\,+\, \frac{d{\sigma}^{L}}{dQ^2\,dp_T^2\,dy}\,(1-3\cos^2\theta)
 \label{hadronwinkel} \\[2mm]
&+& \frac{{d\sigma}^{T}}{dQ^2\,dp_T^2\,dy}\,2\sin^2\theta\cos2\phi
\,+\, \frac{d{\sigma}^{I}}{dQ^2\,dp_T^2\,dy}\,
2\sqrt{2}\sin2\theta\cos\phi \>.
\nonumber
\end{eqnarray}
The unpolarized differential production cross section is denoted by
${\sigma}^{U+L}$ whereas ${\sigma}^{L,T,I}$ characterize the
polarization of the virtual photon, {\it e.g.}, the cross section for
longitudinally polarized virtual photons is denoted by ${\sigma}^L$,
the transverse interference cross section by ${\sigma}^T$, and the
transverse-longitudinal interference cross section by ${\sigma}^I$
(all with respect to the chosen $z$-axis of the lepton pair rest
frame). The hadronic  helicity cross sections
$\frac{d{\sigma}^{\alpha}}{dQ^2\,dp_T^2\,dy}$ in
Eq.~(\ref{hadronwinkel}) are obtained by convoluting the partonic
helicity  cross sections with the parton  densities:
\begin{equation}
\frac{d\sigma^{\alpha}}{ dQ^2\,dp_{T}^{2}\,dy } =
\int\,\,dx_{1}\, dx_{2}\,
f^{h_{1}}(x_{1},\mu_F^{2})\,
f^{h_{2}}(x_{2},\mu_F^{2})\,
\frac{{s}\,d{\hat{\sigma}^\alpha}}{dQ^2\, d{t}\,d{u}} \>.
\label{wqalphahad}
\end{equation}
Introducing the standard angular coefficients~\cite{cs}
\begin{equation}
A_{0}=\frac{2\,\, d\sigma^{L}}{d\sigma^{U+L}} \>, \hspace{1cm}
A_{1}=\frac{2\sqrt{2}\,\, d\sigma^{I}}{d\sigma^{U+L}} \>, \hspace{1cm}
A_{2}=\frac{4 \,\,d\sigma^{T}}{d\sigma^{U+L}} \>, \hspace{1cm}
\label{aintr}
\end{equation}
the angular distribution in Eq.~(\ref{hadronwinkel}) can be
conveniently written
\begin{eqnarray}
\frac {d\sigma}{ dQ^2\,d p_{T}^{2}\,dy\, d\cos\theta \,d\phi}
&=&  \frac{3}{16\pi}\,
\frac{d\sigma^{U+L}}{ dQ^2\,d p_{T}^{2}\,dy}\,\,
           \,\left[  (1+\cos^{2}\theta)
               +\,\, \frac{1}{2}A_{0} \,\,\, (1-3\cos^{2}\theta)
 \right. \label{ang} \\[2mm]
&&   \left.
\hspace{1.5cm} +  \,\,   A_{1}  \,\,\,\sin 2\theta \cos\phi \,\,
\,\, + \,\,   \frac{1}{2}A_{2}  \,\,\,\sin^{2}\theta\cos 2\phi
  \right] \>. \nonumber
\end{eqnarray}
Integrating the angular distribution in Eq.~(\ref{ang}) over the
azimuthal angle $\phi$ yields
\begin{equation}
\frac{d\sigma}{dQ^2\,dp_{T}^{2}\,dy\,d\cos\theta}=C\,
                  (1+\alpha
\cos^{2}\theta) \>,
\label{alphadef}
\end{equation}
where
\begin{equation}
C= \,\frac{3}{8}\,
\frac{d\sigma^{U+L}}{dQ^2\,dp_{T}^{2}\,dy}
\left[1+\frac{A_{0}}{2}\right]\>, \hspace{1cm}
\alpha=\frac{2-3A_{0}}{2+A_{0}} \>.
\label{alphaidef}
\end{equation}
Integrating Eq.~(\ref{ang}) over the polar angle $\theta$ yields
\begin{equation}
\frac{d\sigma}{dQ^2\,dp_{T}^{2}\,dy\,d\phi}=
\,\frac{1}{2\pi}\,
\frac{d\sigma^{U+L}}{dQ^2\,dp_{T}^{2}\,dy} \,
         (1+\beta\cos 2\phi ) \>,
\label{betadef}
\end{equation}
where
\begin{equation}
\beta=\frac{A_{2}}{4} \>. \hspace{1cm}
\label{betaidef}
\end{equation}
By taking  moments with respect to an appropriate product of
trigonometric functions it is possible to disentangle the coefficients
$A_i$. A convenient definition of the moments is~\cite{jimdy}
\begin{equation}
\langle m \rangle =
\frac{\int d\sigma(p_T^{},y,\theta,\phi)\,\, m \,\,
{d\cos\theta}\,{d\phi}}{
\int d\sigma(p_T^{},y,\theta,\phi)\,
\,{d\cos\theta}\,{d\phi}} \>,
\label{moment}
\end{equation}
which leads to the following results:
\begin{eqnarray}
\langle 1 \rangle &=& 1 \>, \\[2mm]
\langle \frac{1}{2}(1-3\cos^2\theta ) \rangle &=&
\frac{3}{20}\,\,\left( A_0-\frac{2}{3}\right) \>, \\[2mm]
\langle \sin 2\theta\cos\phi  \rangle &=& \frac{1}{5} \,\,A_1 \>, \\[2mm]
\langle \sin^2\theta\cos2\phi  \rangle &=&\frac{1}{10}\,\, A_2 \>.
\end{eqnarray}

\subsection{Next-to-leading order cross section}

At ${\cal O}(\alpha_s^2)$ the following partonic  tree level  and
one-loop  processes contribute to  the partonic helicity cross
sections $\frac{{s}\,d{\hat{\sigma}^\alpha}}{dQ^2\, d{t}\,d{u}}$  in
Eq.~(\ref{wqalphahad}).
\begin{equation}
    \begin{array}{llll}
\mbox{tree level contributions}:   &
         q     +   \bar{q}  & \rightarrow &   \gamma^*  +   g   +    g    \>,\\
      &  q     +   \bar{q}  & \rightarrow &   \gamma^*  +   q   + \bar{q} \>,\\
      &  q     +   g        & \rightarrow &   \gamma^*  +   q   +    g    \>,\\
      &  q     +   q        & \rightarrow &   \gamma^*  +   q   +    q    \>,\\
      &  g     +   g        & \rightarrow &   \gamma^*  +   q   + \bar{q} \>,\\
    \end{array}
  \label{tree}
\end{equation}
\begin{equation}
    \begin{array}{llll}
\mbox{one-loop contributions}:   &
         q     +   \bar{q}  & \rightarrow & \gamma^*  + g \>,\\
      &  q     +   g        & \rightarrow & \gamma^*  +   q \>.\\
    \end{array}
  \label{loop}
\end{equation}
The second-order contributions in Eq.~(\ref{loop}) come from the
interference of the one-loop diagrams with the leading-order diagrams.
Let us briefly sketch the technical ingredients that go into our
calculation (for more details, see Ref.~\cite{npb}). For the ${\cal
O}(\alpha_s^2)$ tree level contributions in Eq.~({\ref{tree}}) we
introduce the variable
\begin{equation}
s_{2}=(k_{1}+k_{2})^{2}=(p_{1}+p_{2}-Q)^{2}=s+t+u-Q^2\>,
\label{s2def}
\end{equation}
in addition to $s,t,u$ defined in Eq.~(\ref{kleinmandeldef}); $s_{2}$
is the invariant mass of the system recoiling against the virtual
photon.

To obtain the $p_{T}^{}$ distribution of the virtual photon, the
${\cal O}(\alpha_s^2)$ tree level diagrams  must to be integrated over
the phase space of the two final state partons with the $p_T^{}$ of
the virtual photon held fixed. The integration over the recoiling
partons is most easily performed in the $(k_{1}k_{2})$ center of mass
system  by integrating  over the solid angle $d\Omega_{k_{1}k_{2}}$.
Since all partons are massless, collinear divergencies appear  after
integrating over   $d\Omega_{k_{1}k_{2}}$. Soft gluon  singularities
show up as poles in the variable $s_{2}$. A finite next-to-leading
order (NLO) partonic helicity cross section is derived in the
following manner.

\begin{itemize}
\item
Infrared and collinear divergencies associated with final state
partons cancel among loop and tree diagrams.
\item
Collinear initial state divergencies are absorbed into the parton
densities, {\it i.e.}, they are removed by renormalizing the parton
densities, which introduces  a factorization scale dependence into the
parton densities $f(x,\mu_F^2)$.
\item
Ultraviolet divergencies are removed by $\overline{MS}$  (Modified
Minimal Subtraction~\cite{MSBAR}) renormalization, which introduces a
renormalization scale dependence into the strong coupling  constant
$\alpha_s(\mu_R^2)$.
\end{itemize}

Following Ref.~\cite{npb}, we introduce the following notation to list
the  partonic helicity cross sections
$\frac{s\,d\hat{\sigma}^{\alpha}_{ab}}{dQ^2\, dt\,du}$ in
Eq.~(\ref{wqalphahad})
$[\alpha\in \{U+L,L,T,I\}]$.\\
${\cal O}(\alpha_s)$ Born contributions:
\begin{equation}
\frac{s\,d\hat{\sigma}^{\alpha,{\rm Born}}_{ab}}{dQ^2\, dt\,du} =
\frac{K_{ab}^{\gamma^*}}{s}\,\,\alpha_s\,\,\delta(s+t+u-Q^2)\,\,
T^{\alpha}_{ab}(\mbox{B}) \>,
\label{borntilde}
\end{equation}
${\cal O}(\alpha_s^2)$ virtual corrections:
\begin{equation}
\frac{s\,d\hat{\sigma}^{\alpha,{\rm virt}}_{ab}}{dQ^2\, dt\,du} =
\frac{K_{ab}^{\gamma^*}}{s}\,\,\frac{\alpha_{s}^{2}}{2\pi}
\,\,\delta(s+t+u-Q^2)\,\,V(\varepsilon)\,\,T^{\alpha}_{ab}(\mbox{V}) \>,
\label{virtualtilde}
\end{equation}
${\cal O}(\alpha_s^2)$ tree graph corrections:
\begin{equation}
\frac{s\,d\hat{\sigma}^{\alpha,{\rm tree}}_{ab}}{dQ^2\, dt\,du} =
\frac{K_{ab}^{\gamma^*}}{s}\,\,\frac{\alpha_{s}^{2}}{2\pi}
\,\,V(\varepsilon)\,\,T^{\alpha}_{ab}(\mbox{T}) \>.
\label{baumtilde}
\end{equation}
The subscript $ab$ stands for the initial state parton pair, {\it
i.e.},  $a$ and $b$ denote a  quark, antiquark, or gluon. The initial
state collinear singularities  have been factorized from
Eq.~(\ref{baumtilde}) at a scale $\mu_F^{2}$. The constants
$K_{ab}^{\gamma^*}$ and $V(\varepsilon)$ are  given by (we work in
$n=4-2\varepsilon$ dimensions)
\begin{eqnarray}
K^{\gamma^*}_{qq}&=&\frac{16\pi}{3}\,\frac{\alpha^{2}}
{8\pi Q^2}
\frac{\,\mbox{$C_{F}$}}{\,\mbox{$N_{C}$}\,}
\frac{(1-\varepsilon)}{\Gamma(1-\varepsilon)}
\left(\frac{4\pi\mu^{2}}{Q^2}\right)^{\varepsilon}
\left(\frac{sQ^2}{ut}\right)^{\varepsilon} \>,    \label{kabdef}
\\[2mm]
K^{\gamma^*}_{qg}&=&\frac{K^{\gamma^*}_{qq}}{2\,\mbox{$C_{F}$}
\,(1-\varepsilon)} \>,
\\[2mm]
K^{\gamma^*}_{gg}&=&\frac{K^{\gamma^*}_{qq}}{4 C_{F}^{2}\,
(1-\varepsilon)^{2}} \>,
\\[2mm]
V(\varepsilon) &=& \frac{\Gamma(1-\varepsilon)}{\Gamma(1-2\varepsilon)}
\left(\frac{4\pi\mu^{2}}{Q^2}\right)^{\varepsilon} \>.
\label{vepsidef}
\end{eqnarray}
$T_{ab}^{\alpha}(\mbox{B,V,T})$ are the partonic helicity matrix
elements for the Born, virtual, and tree contributions. The explicit
form of the partonic helicity matrix elements depends on the choice of
the $z$-axis in the $\gamma^*$ rest frame. Covariant projection cross
sections $d\hat{\sigma}^{\beta}\,\,\,
[\beta\in\{U+L,L_1,L_2,L_{12}\}]$ for $W$ boson production have been
calculated to ${\cal O}(\alpha_s^2)$
in in Ref.~\cite{npb}. From these results  one can
obtain the relevant helicity  cross sections in
Eqs.~(\ref{borntilde})--(\ref{baumtilde}) for any given  $\gamma^*$
rest frame [see Eqs.~(\ref{csmatrix}) and (\ref{gjmatrix})].  However,
there are some further contributions in the case of $\gamma^*$
production, which do not contribute to $W$ boson production because of
charge conservation. The diagrams can be found in Figs.~8 and 9 of
Ref.~\cite{npb}. The remaining analytical projections for these
contributions (after integration over $d\Omega_{k_1k_2}$)  are listed
here in an appendix. The folding of the NLO parton level cross
sections with the respective parton densities is straightforward and
we do not list the combinations here.

\subsection{The $z$-axis in the lepton pair rest frame}

Before we present numerical results for the angular distributions, it
is necessary to discuss the choice of the $z$-axis in the lepton-pair
rest frame. We will discuss two different choices:  the Collins-Soper
(SC) frame~\cite{cs} and the Gottfried-Jackson (GJ)
frame~\cite{gottfried}. In the CS frame the $z$-axis bisects the angle
between ${\vec{P}_{1}}$ and ${-\vec{P}_{2}}$:
\begin{equation}
          \begin{array}{ll}
&{\vec{P}_{1}} = E_{1}\,(\sin\gamma_{CS}^{},0, \cos\gamma_{CS}^{}) \>, \\[1mm]
&{\vec{P}_{2}} = E_{2}\,(\sin\gamma_{CS}^{},0,-\cos\gamma_{CS}^{}) \>,
          \end{array}
\end{equation}
with
\begin{eqnarray}
\cos\gamma_{CS}^{} &=& \left( \frac{Q^2 S}{(T-Q^2)(U-Q^2)} \right)^{1/2}
                    =  \left( \frac{Q^2}{Q^2+p_T^2} \right)^{1/2} \>,
\label{winkelcsdef} \\[1mm]
\sin\gamma_{CS}^{} &=& -\sqrt{1-\cos^2\gamma_{CS}^{}} \>,\\[1mm]
E_1&=&\frac{Q^2-T}{2\sqrt{Q^2}}\>,
\hspace{1cm}
E_2=\frac{Q^2-U}{2\sqrt{Q^2}}\>.
\end{eqnarray}
In the GJ frame the $z$-axis is chosen parallel to the beam axis:
\begin{equation}
          \begin{array}{ll}
          & {\vec{P}_{1}} = E_{1} \,\,\,
                            (0,0,1)\>,\\[1mm]
          & {\vec{P}_{2}} = E_{2}\,\,\,
                            (\sin\gamma_{GJ},0,\cos\gamma_{GJ})\>,
          \end{array}
\label{gjdef}
\end{equation}
with
\begin{eqnarray}
\cos\gamma_{GJ}^{} &=&
 {1-\frac{2Q^2 S}{(T-Q^2)(U-Q^2)}}
= \frac{p_{T}^2-Q^2}{p_{T}^2+Q^2} \>, \\[1mm]
\sin\gamma_{GJ}^{} &=& -\sqrt{1-\cos^2\gamma_{GJ}} \>, \\[1mm]
E_{1} &=& {Q^2-T \over 2Q} \>, \hspace{1cm} E_{2}={Q^2-U \over 2Q} \>.
\end{eqnarray}
Note that the CS and GJ frames are related by a rotation about the
$y$-axis. In the laboratory frame, the $z$-direction is defined by the
proton momentum and the $x$-direction is defined by the transverse
momentum of the virtual photon.

\section{Numerical Results}

In this section numerical results are presented for high $p_T^{}$
production and leptonic decay of a virtual photon at the Tevatron
collider center of mass energy ($\sqrt{S} = 1.8$~TeV). The numerical
results have been obtained using the MRS set $A$~\cite{mrs} parton
distribution functions  with  $\Lambda_{\overline{MS}}^{(4)}=230$~MeV.
For our NLO predictions, we use the two-loop formula for $\alpha_s$
with five favors. If not stated otherwise, the renormalization scale
$\mu_R^2$ and the factorization scale $\mu_F^2$ in Eq.~(\ref{wqhad})
have been taken to be $\mu_R^2 = \mu_F^2 = [Q^2+p_T^2(\gamma^*)]/2$,
where $Q$ and $p_T^{}(\gamma^*)$ are the invariant mass and transverse
momentum, respectively, of the virtual photon. We work in the
$\overline{MS}$ factorization scheme.

We begin with numerical results for the coefficients $A_i$, $\alpha$,
and $\beta$ in Eqs.~(\ref{ang}), (\ref{alphadef}), and
(\ref{betadef}).
Figure~1a shows the coefficients $A_0,A_1$, and
$A_2$ in the CS frame as functions of $p_T^{}(\gamma^*)$
for an
invariant mass $Q$ of the photon fixed to $Q = 10$~GeV.
The reason for choosing this fairly large invariant mass is to minimize
the effect of the acceptance cuts
which will later be imposed on the decay leptons [see below].
The dotted
lines are leading order (LO) [${\cal O}(\alpha_s^1)$]  predictions and
the solid lines are next-to-leading order [${\cal O}(\alpha_s^2)$]
predictions. The coefficients $A_0$ and $A_2$ are increasing functions
of $p_T^{}(\gamma^*)$ and the deviations from the zero order  $[{\cal
O}(\alpha_s^0)]$ expectation [$A_0=A_2=0$ at $p_T^{}(\gamma^*)=0$] are
quite large, even at modest values of $p_T^{}(\gamma^*)$. It has been
noted in  Ref.~\cite{lo} that the coefficients $A_0$ and $A_2$ are
exactly equal at LO (dotted line). This is no longer true at NLO, but
the corrections  are fairly small, especially the corrections to
$A_0$. The ${\cal O}(\alpha_s^2)$ corrections to $A_2$ are negative
and about 20\% the size of the LO result. Note that $A_0$ originates
from the longitudinal polarization  of the virtual photon, whereas
$A_2$ receives contributions from a transversely polarized virtual
photon [all with respect to the $z$-axis of the chosen lepton pair
rest frame]. The deviation of $A_1$
(interference of longitudinal and transverse virtual photon
polarizations) from zero is  small in the CS frame, even at large
values of $p_T^{}(\gamma^*)$.

Figure~1b shows numerical results for the coefficients $\alpha$ and
$\beta$ [see Eqs.~(\ref{alphadef})  and (\ref{betadef})] as a function
of $p_T^{}(\gamma^*)$. The coefficients are again very sensitive to
the transverse momentum  of the virtual photon and the deviations from
the zero order predictions ($\alpha=1$ and $\beta=0$) are again large.
The NLO corrections to $\alpha$ are small over the whole range of
$p_T^{}(\gamma^*)$.

Numerical results for the coefficients $A_i$, $\alpha$, and $\beta$
in the GJ frame are shown in Fig.~2. The corrections to $A_0$ and
$\alpha$ are larger than in the CS frame, however, in both frames the
corrections do not dramatically change the LO results. This is because
the coefficients $A_i$ are ratios of  helicity cross sections [see
Eq.~(\ref{aintr})] and the large QCD corrections in the individual
helicity cross sections tend to cancel in the ratios. This can be seen
from Fig.~3, where the $K$-factors for the helicity cross sections
$\sigma^{U+L},\sigma^{L}$, and $\sigma^T$ are shown as functions of
$p_T^{}(\gamma^*)$ in the CS frame (Fig.~3a) and  GJ frame (Fig.~3b).
The $K$-factor is defined as  the ratio of the NLO [${\cal
O}(\alpha_s^2)$] differential cross section  to the LO [${\cal
O}(\alpha_s^1)$] differential cross section. Results are shown in
Fig.~3 for two different choices of the scale  $\mu^2 = \mu^2_F =
\mu^2_R$: the upper curves correspond to $\mu^2 = 1/4\,
[p_T^2(\gamma^*) + Q^2]$ and the lower curves correspond to $\mu^2 =
4\, [p_T^2(\gamma^*) + Q^2]$. The invariant mass of the virtual photon
has again been fixed to $Q = 10$~GeV. The $K$-factor for $\sigma^{U+L}$
and $\sigma^{L}$  ranges from 0.9 to 1.6  in the CS frame depending on
$p_T^{}(\gamma^*)$ and the choice of the renormalization and
factorization scale. However, the $K$-factor is almost the same for
$\sigma^{U+L}$ and $\sigma^{L}$, thus $A_0$, which is proportional to
the ratio $\sigma^L / \sigma^{U+L}$, is not effected by the
corrections. In the GJ frame, the  $K$-factors for $\sigma^L$ and
$\sigma^{U+L}$ differ more, and thus the corrections to $A_0$ are
larger in the GJ frame (see Fig.~2b). The $K$-factors  for $\sigma^T$
are particularly different from the $K$-factors for $\sigma^{U+L}$,
which explains the large deviation of $A_2$ from the LO result
$A_0=A_2$. Note that the  large $K$-factors are due  to large
logarithms [like $\ln(s/Q^2)$] in the NLO matrix elements~\cite{npb}.
The $K$-factors decrease with increasing invariant mass of the virtual
photon.

To give a feeling for the numerical contributions from the different
partonic subprocesses, Fig.~4 shows the fractional contributions
of the partonic subprocesses to
$\sigma^{U+L},\sigma^{L}$, and $\sigma^{T}$ for $\gamma^*$ production
as a function of $p_T^{}(\gamma^*)$ for $Q=10$~GeV. The curves labeled
$A$ and $B$ correspond to the LO results from the subprocesses $q \bar
q \to \gamma^* g$ and $q g \to \gamma^* q$, respectively.  The curves
labeled $C$ through $G$  correspond to the  ${\cal O}(\alpha_s^2)$
contributions from the subprocesses listed in Eqs.~(\ref{tree}) and
(\ref{loop}). Note that the NLO contribution from the $qg$ initiated
subprocess (curve $E$) is more important than the LO contribution from
the $q\bar{q}$  subprocess (curve A) at large values of
$p_T^{}(\gamma^*)$. The other ${\cal O}(\alpha_s^2)$ subprocesses are
fairly small and  even give negative contributions at small values of
$p_T^{}(\gamma^*)$. This is due to the factorization of the collinear
initial state singularities  which dominate these subprocesses  at low
values of  $p_T^{}(\gamma^*)$.  Although the total contributions from
the ${\cal O}(\alpha_s^2)$ subprocesses  are large, they do not have a
dramatic effect on the decay lepton distribution (see Figs.~1 and 2),
{\it i.e.}, the  polarization of the virtual photon as a function of
its transverse momentum is almost unchanged by higher order
corrections.

We want to point out that direct measurements of the coefficients
$A_i$ for virtual photon production  in pion-nucleon scattering  at
$\sqrt{S} \approx 19$ and $23$~GeV are not in agreement with the LO
QCD predictions~\cite{na10,cip}. The NLO corrections  to the
coefficients are very small in this case (see Fig.~7 in
Ref.~\cite{nachtmann2}) and the results in Refs.~\cite{na10} and~\cite{cip}
cannot be explained by standard Drell-Yan production supplemented with
QCD corrections.

We now turn our attention to the $\cos\theta$
and $\phi$ distributions of the decay leptons. Since the effects of
the NLO corrections are small (in particular for the $\cos\theta$
distribution in the CS frame; see the coefficients $A_0$ and $\alpha$
in Fig.~1), it is sufficient to use LO matrix elements in our Monte
Carlo study of the lepton decay distributions. Results will be shown
for the $\phi$ and $\cos\theta$ distributions of  leptons originating
from the decay of a virtual photon produced  with finite transverse
momentum in $p \bar p$ collisions at the Tevatron center of mass
energy. To demonstrate the effects of cuts, results are shown first
without cuts and then with typical acceptance cuts imposed on the
leptons. These cuts are necessary due to the finite acceptance of the
detector.

Measurement uncertainties, due to the finite energy resolution of the
detector, have been simulated in our calculation by Gaussian smearing
of the lepton four-momentum vectors with standard deviation $\sigma$.
The numerical results presented here were made using $\sigma$ values
based on the CDF specifications~\cite{SMEARING}. The energy resolution
smearing has a negligible effect on the $\phi$ and $\cos\theta$
distributions.

Figure~5 shows the normalized $\phi$ and $\cos\theta$ distributions of
the  decay leptons from $\gamma^*$ production  for three bins in the
transverse momentum of the virtual photon. The invariant mass of the
photon has been integrated over the range
$10\ {\rm GeV} < Q < 12\ {\rm GeV}$.
No cuts or
smearing have been applied to the results in the figure. The  curves
in  Fig.~5 can be obtained  from $\alpha$ and $\beta$ in Fig.~1b.
[Note, however, that Fig.~1 shows the coefficients $\alpha$ and
$\beta$ for a fixed value of $Q=10$~GeV.] For example,  since $\beta$
is an increasing function of $p_T^{}(\gamma^*)$, the amplitude of the
$\phi$ distribution increases with $p_T^{}(\gamma^*)$. Likewise,
$\alpha$ is a decreasing function of $p_T^{}(\gamma^*)$, starting out
positive and  ending up negative, thus the curvature of the
$\cos\theta$ distribution in Fig.~5b is positive for
the two lowest
$p_T^{}(\gamma^*)$ bins and is nearly zero for the
$p_T^{}(\gamma^*) > 6$~GeV bin.
Note that if the virtual photon
was to decay isotropically, the $\phi$ and $\cos\theta$ distributions
would both be flat.

The effect of acceptance cuts on the angular distributions is
illustrated in  Fig.~6 which shows the $\phi$ and $\cos\theta$
distributions of the decay leptons for the same bins in
$p_T^{}(\gamma^*)$ as in  Fig.~5, but now with energy resolution
smearing and the cuts
\begin{equation}
p_T^{}(\ell) > 2\ \mbox{GeV},
\hspace{20mm}
|y(\ell)| < 2.5 \>,
\hspace{20mm}
|y(\gamma^*)|<1.0 \>.
\label{loosecuts}
\end{equation}
The cuts on the leptons are necessary due to the finite acceptance of the
detector.  The photon rapidity cut has been imposed because we find that
polarization effects are highlighted when the virtual photon is in the central
rapidity region.
The cuts on the leptons, in particular the $p_T^{}(\ell)$ cut, have
a dramatic effect on the shapes of the distributions.
The shapes of the angular distributions are now governed by the kinematics of
the surviving events.
Only 20\% of the
events pass these cuts.
The cuts in Eq.~(\ref{loosecuts}),  which are applied in
the laboratory frame, introduce a strong $\phi$ dependence.
The ``kinematical'' $\phi$ distribution in Fig.~6a is  very different from
the ``dynamical'' $\phi$ distribution in Fig.~5a, in fact,
the peaks and valleys are interchanged in the two distributions.
The only remaining
vestiges of the polarization  effects in the $\phi$ distribution are
the dips in the high $p_T^{}(\gamma^*)$ curve (solid line) at $\phi =
90^\circ$ and $270^\circ$. The $\cos\theta$ distributions with cuts in
Fig.~6b are also  different from the corresponding  results
without cuts  in Fig.~5b, in particular for $|\cos\theta| \gtrsim 0.5$.
However, for $|\cos\theta| < 0.5$ and small $p_T^{}(\gamma^*)$
(dotted and dashed curves in Fig.~6b) the polarization effects
in the $\cos\theta$ distributions are still visible.
As stated earlier, the effect of energy
resolution smearing is  negligible; the drastic changes in the shapes
of the distributions are due to the cuts, especially the
$p_T^{}(\ell)$ cut.
The polarization effects diminish as the invariant mass of
the virtual photon decreases.
The LO cross section (summed over $\ell = e,\mu$)
for the three $p_T^{}(\gamma^*)$ bins,
2~GeV  $<\, p_T^{}(\gamma^*)\, < 4$~GeV,
4~GeV  $<\, p_T^{}(\gamma^*)\, < 6$~GeV,
$p_T^{}(\gamma^*) > 6$~GeV,
are 50~pb, 20~pb, 26~pb, respectively.
However, these numbers should be multiplied by  the $K$-factor for
$\sigma^{U+L}$ as shown in Fig.~3. Furthermore, soft gluon
resummation effects
will also be important for the production cross section at low
$p_T^{}(\gamma^*)$.

The cuts in Eq.~(\ref{loosecuts}) are actually quite weak. Figure~7
shows the $\phi$ and $\cos\theta$ distributions with the stronger and
more realistic cuts
\begin{equation}
p_T^{}(\ell) > 5\ \mbox{GeV},
\hspace{20mm}
|y(\ell)| < 1.0 \>,
\hspace{20mm}
|y(\gamma^*)|<1.0 \>,
\label{tightcuts}
\end{equation}
for the same three $p_T^{}(\gamma^*)$ bins as in Fig.~5.
The curves are now very different from the curves in Fig.~5 and no
traces of polarization effects are left in Fig.~7.

We have also analyzed  the effect of the cuts by using  the
correct matrix element for $\gamma^*$  production, but with isotropic
decay of the virtual photon, {\it i.e.}, neglecting spin correlations between
$\gamma^*$ production and decay. The angular distributions in this
case are  similar to the ones shown in Figs.~6 and 7 for the full
matrix element;  the remnant polarization effects discussed in Fig.~6
are of course absent.

In Fig.~8 we show ratios of the $\phi$ and $\cos\theta$ distributions
for the same bins in $p_T^{}(\gamma^*)$ as in Fig.~5;  the
distribution with full polarization has been divided by the corresponding
distribution obtained with isotropic decay of the virtual photon.
Cuts and smearing are included in both cases. The large effects from
the cuts are expected to almost cancel in this ratio. In
fact, we nearly recover the $\phi$ and $\cos\theta$ dependence of Fig.~5
which contains no cuts.

Fig.~9 shows the ratio of the $\phi$ and $\cos\theta$ distributions
with full polarization to  the corresponding distributions obtained
with isotropic leptonic decay for the virtual photon for the cuts in
Eq.~(\ref{tightcuts})
for the high $p_T^{}(\gamma^*)$ bin, {\it i.e.}, $p_T^{}(\gamma^*)>6$ GeV.
The kinematical effects in the two low
$p_T^{}(\gamma^*)$ bins are very large and the cuts remove all events
around $\phi=0^\circ, 180^\circ, 360^\circ$ as well as for
large $\cos\theta$ values [see Fig.~7], thus it is impractical to form
ratios for the two low $p_T^{}(\gamma^*)$ bins.
However,
the ratios for $p_T^{}(\gamma^*) > 6$~GeV  [see Fig. 9] once again
contain most of the polarization dependence seen in the solid curves
of Fig.~5.
The
additional dips in the $\phi$ distribution
at $\phi=0^\circ,180^\circ$, and
$360^\circ$ in Fig.~9a are due to the kinematical
cuts.
Thus even in the presence of large acceptance cuts
it may be possible to highlight the
polarization effects in the experimental results
by dividing the
experimental distributions by the corresponding Monte Carlo
distributions obtained using isotropic  $\gamma^*$ decay.

\section{SUMMARY}

The polar and azimuthal angular distributions of the lepton pair
arising from the decay of a virtual photon produced at high transverse
momentum in hadronic collisions have been discussed.  In the absence of
cuts on the final state leptons, the general structure of the lepton
angular distribution in the virtual photon rest frame is determined by
the polarization of the virtual photon.   In perturbative QCD, the
structure is described by four helicity cross sections, which are
functions of the transverse momentum and rapidity of the virtual
photon.   We have calculated to ${\cal O}(\alpha_s^2)$
the angular coefficients which govern the lepton angular distributions
and find that the
corrections are relatively  small in both the CS and GJ frames,
especially in the CS frame. This is because the angular coefficients
$A_i$ are ratios of helicity cross sections and the large QCD
corrections in the individual helicity cross sections tend to cancel
in the ratios.

We have also studied the angular distributions of the leptonic decay
products of a high $p_T^{}$ virtual photon when acceptance cuts and
energy resolution smearing are applied to the leptons.   When
acceptance cuts are imposed on the leptons, the shapes of the lepton
angular distributions are dominated by kinematic effects
and the residual
dynamical effects from the virtual photon polarization are small.
The kinematic effects become more dominate as the cuts become more
stringent and as the invariant mass of the photon decreases.
Energy resolution smearing has a negligible effect  on the
angular distributions.

Polarization effects can be maximized by minimizing the cuts on the
decay leptons, however,  this strategy is severly limited since cuts
are needed due to the  finite acceptance of a detector.
Polarization effects are also more pronounced when the
virtual photon is in the central rapidity region.
Alternatively, it may be possible to highlight virtual photon
polarization effects by ``dividing out'' the kinematic effects, {\it
i.e.}, if the histogrammed data is divided by the theoretical result
for isotropic virtual photon decay, the resulting ratio is more
sensitive to polarization effects.

\begin{center}
{\bf Acknowledgements}
\end{center}

Helpful discussions with O.~Nachtmann,  M.~I.~Martin, and  D.~Wood  are
gratefully acknowledged. This work is supported in part by the U.S.
Department of Energy under contract Nos. DE-AC02-76ER00881 and
DE-FG03-91ER40674, and by the University of Wisconsin Research
Committee with funds granted by the Wisconsin Alumni Research
Foundation.

\newpage
\begin{center}
{\bf FIGURE CAPTIONS}
\end{center}
\begin{itemize}
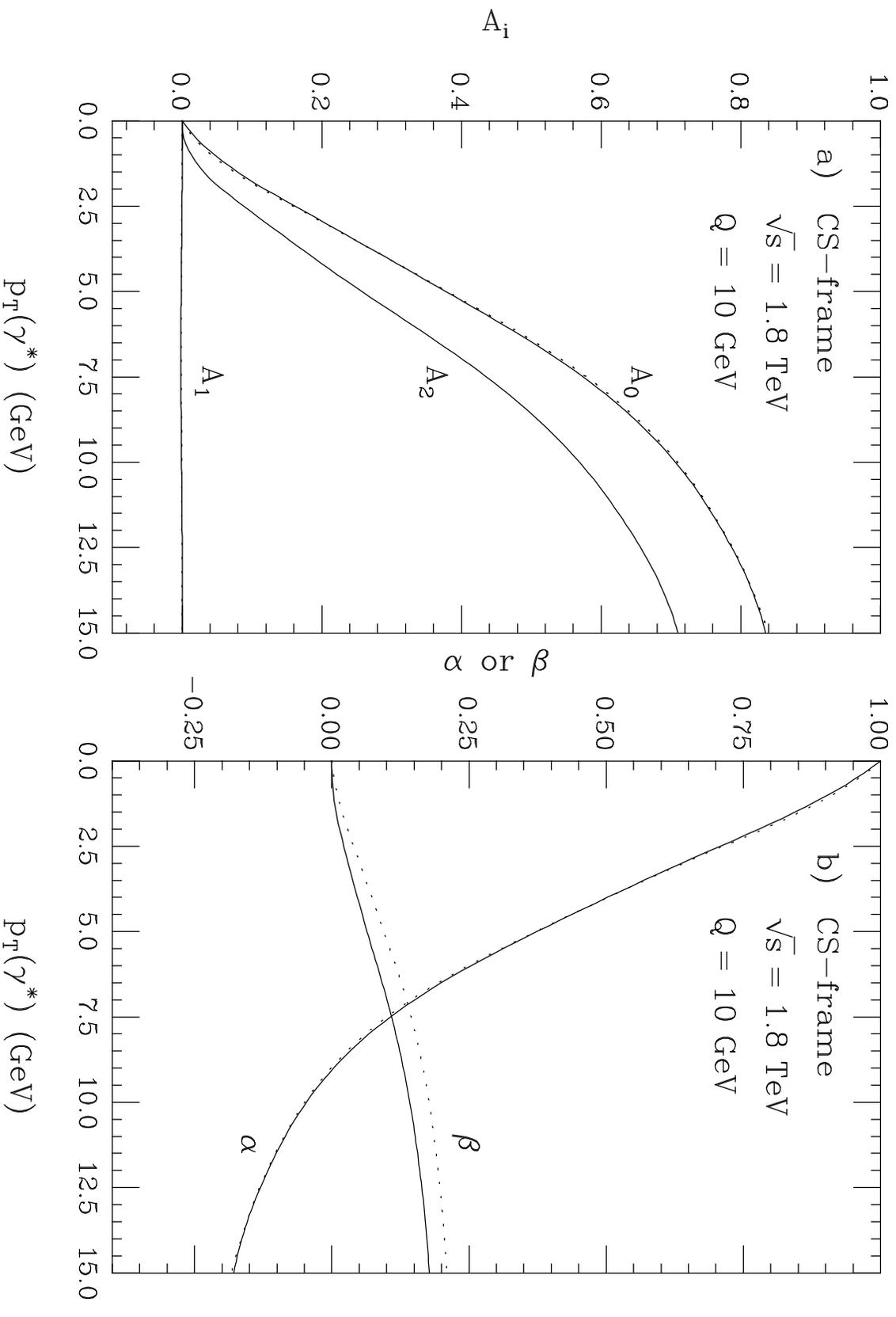
\item[{\bf Fig. 1}]
Angular coefficients for $\gamma^*$ production and decay in the CS
frame as a function of the transverse momentum $p_T^{}(\gamma^*)$ for
$Q = 10$~GeV and $\sqrt{S} = 1.8$~TeV. Part a) shows the angular
coefficients  $A_0, A_1$, and  $A_2$ and  part b) shows the  angular
coefficients $\alpha$ and $\beta$. The dotted lines are LO predictions
($A_0=A_2$ at LO) and the solid lines are NLO predictions.  No cuts or
smearing have been applied.

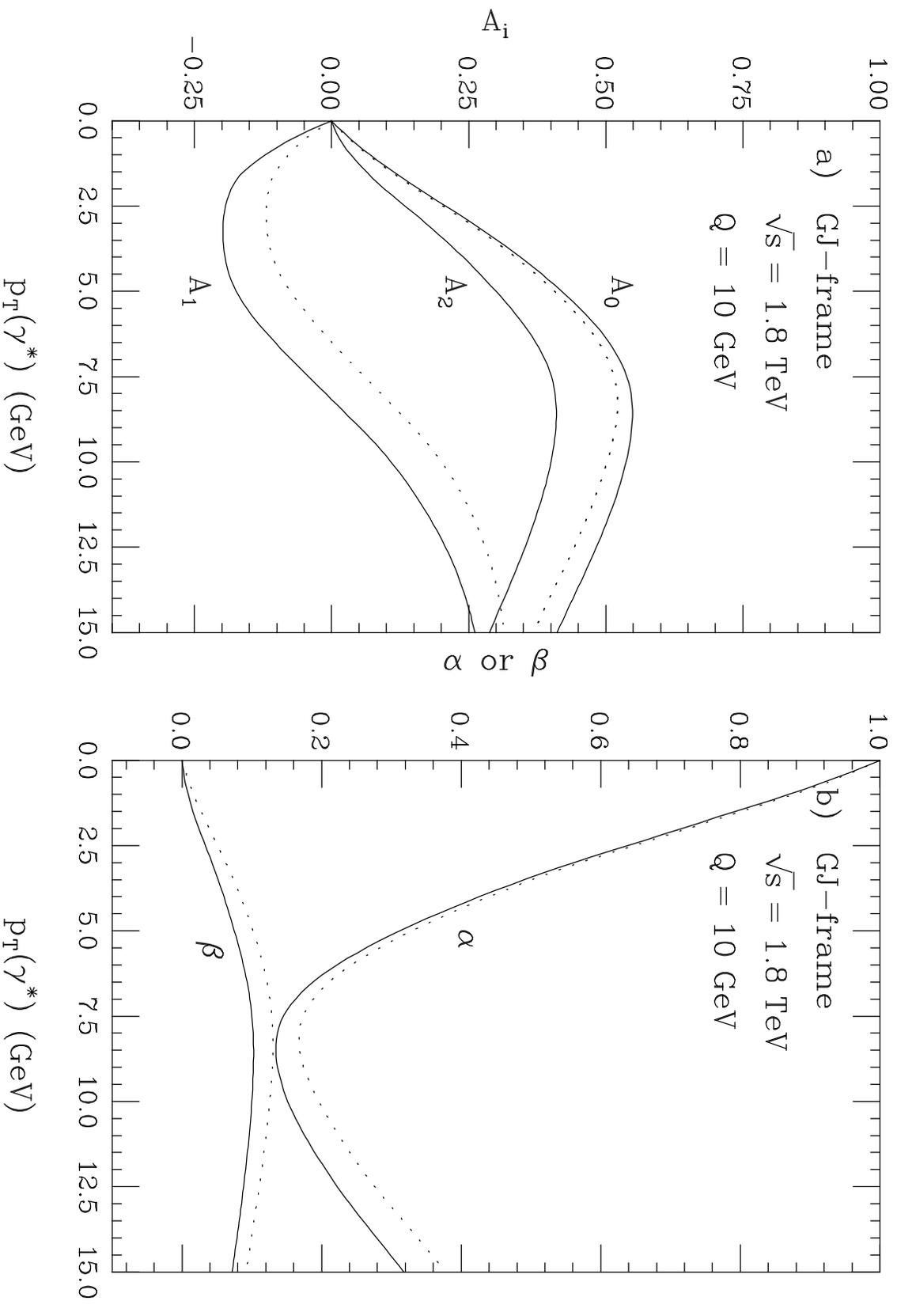
\item[{\bf Fig. 2}]
Same as Fig.~1 but for the GJ frame.

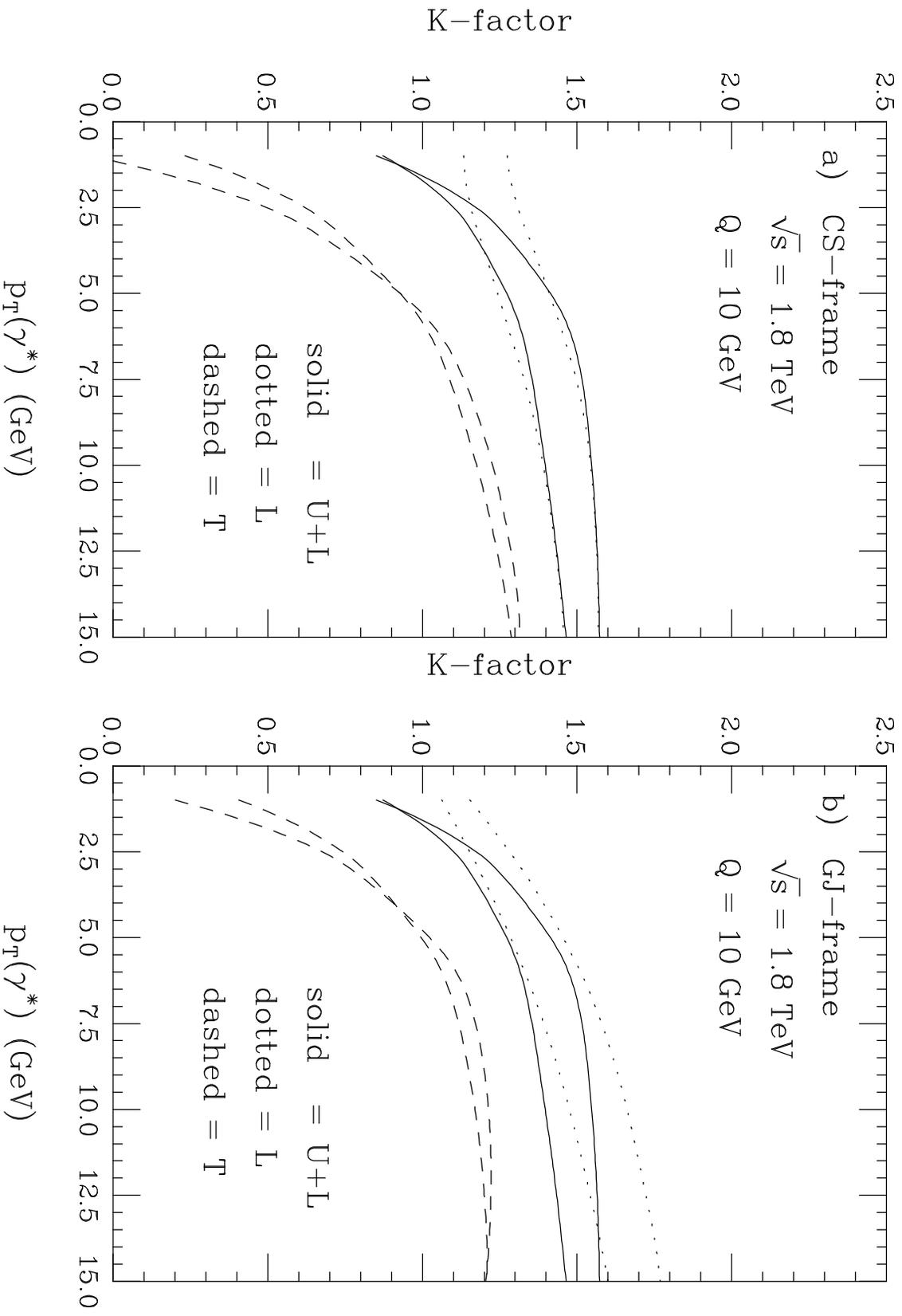
\item[{\bf Fig. 3}]
$K$-factors for the helicity cross sections  $d\sigma^{U+L},
d\sigma^{L}$, and $d\sigma^{T}$ for $\gamma^*$ production and decay as
a function of the transverse momentum $p_T^{}(\gamma^*)$ for $Q=10$~GeV
and $\sqrt{S}=1.8$~TeV. Parts a) and b) are for the CS and GJ frames,
respectively. Results are shown for two different choices of the scale
$\mu^2=\mu_F^2=\mu_R^2$: the upper lines are for $\mu^2= 1/4
\,[p_T^2(\gamma^*) + Q^2]$ and the lower lines are for $\mu^2= 4
\,[p_T^2(\gamma^*) +Q^2]$.

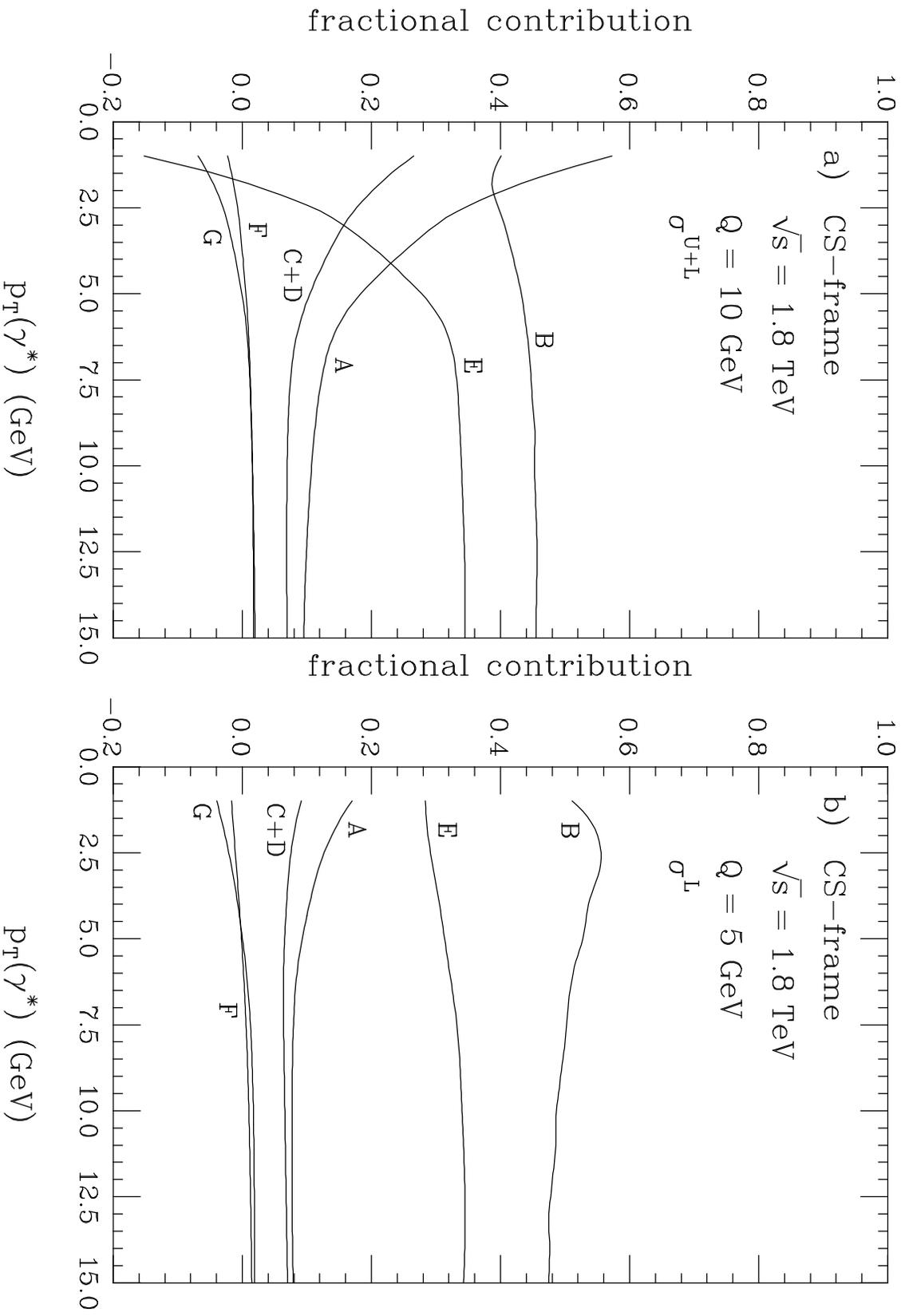
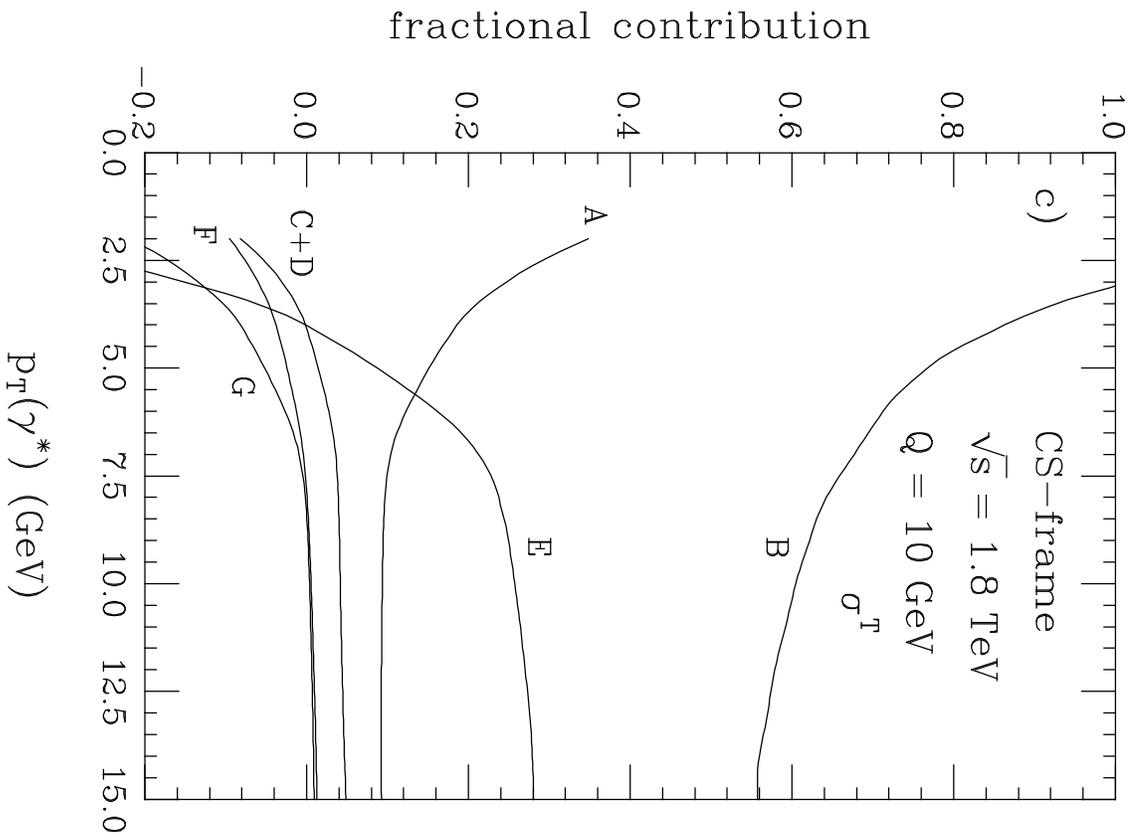
\item[{\bf Fig. 4}]
a) Fractional contributions to $d\sigma^{U+L}$   for $\gamma^*$
production in the CS frame as a function of  the transverse momentum
$p_T^{}(\gamma^*)$ for $Q=10$~GeV, $\sqrt{S}=1.8$~TeV, and $\mu^2 = 1/2
\, [p_T^2(\gamma^*) + Q^2]$. The curves are labeled according to the
contributing subprocesses.
\begin{quotation}
 \noindent
 \vspace{-2mm}
The LO subprocesses are:\\
(A) $q \bar{q} \rightarrow \gamma^* g$ and
(B) $q      g  \rightarrow \gamma^* q$.\\
The NLO subprocesses are:\\
(C+D) $(q\bar{q} \rightarrow \gamma^* g) +
       (q\bar{q} \rightarrow \gamma^* gg) +
       (q\bar{q} \rightarrow \gamma^* q \bar q)$,\\
(E)\hspace{6mm}
  $({q} g \rightarrow \gamma^* {q}) +
   ({q} g \rightarrow \gamma^* {q} g)$, \\
(F)\hspace{8mm} ${q}{q} \rightarrow \gamma^* {q}{q}$,\\
(G)\hspace{6mm} $gg \rightarrow \gamma^* q \bar{q}$.
   \end{quotation}
b) Same as a) but for $d\sigma^{L}$.\\
c) Same as a) but for $d\sigma^{T}$.

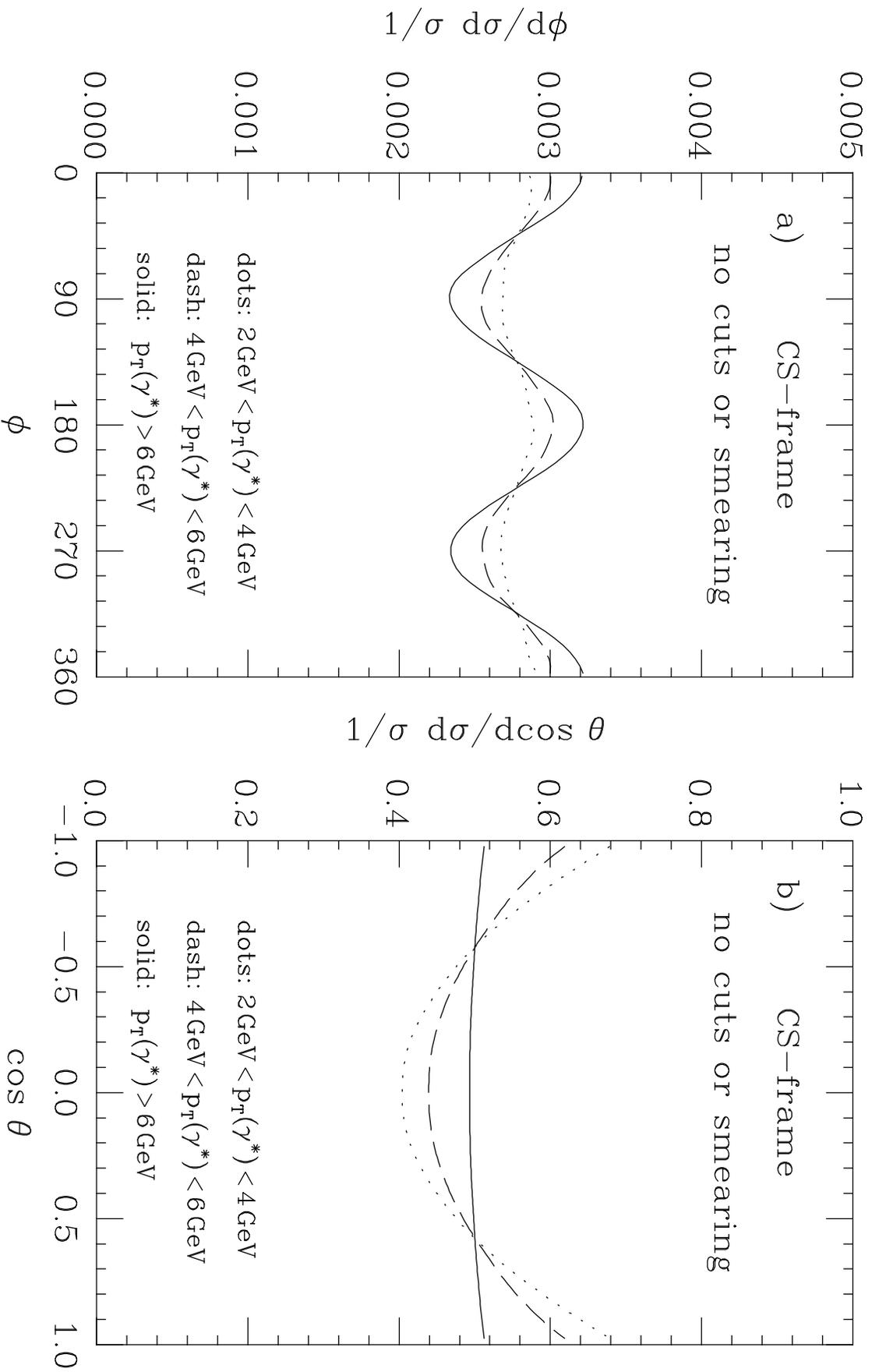
\item[{\bf Fig.~5}]
a) Normalized $\phi$ and b) normalized $\cos\theta$ distributions
of the leptons from $\gamma^*$ decay
in the CS frame with $10\ {\rm GeV} < Q < 12\ {\rm GeV}$.
Results are shown for three bins in $p_T^{}(\gamma^*)$:\\
2~GeV  $<\, p_T^{}(\gamma^*)\, < 4$~GeV (dotted),\\
4~GeV  $<\, p_T^{}(\gamma^*)\, < 6$~GeV (dashed),\\
$p_T^{}(\gamma^*)\, > 6$~GeV (solid). \\
No cuts or smearing have been applied to the decay leptons.

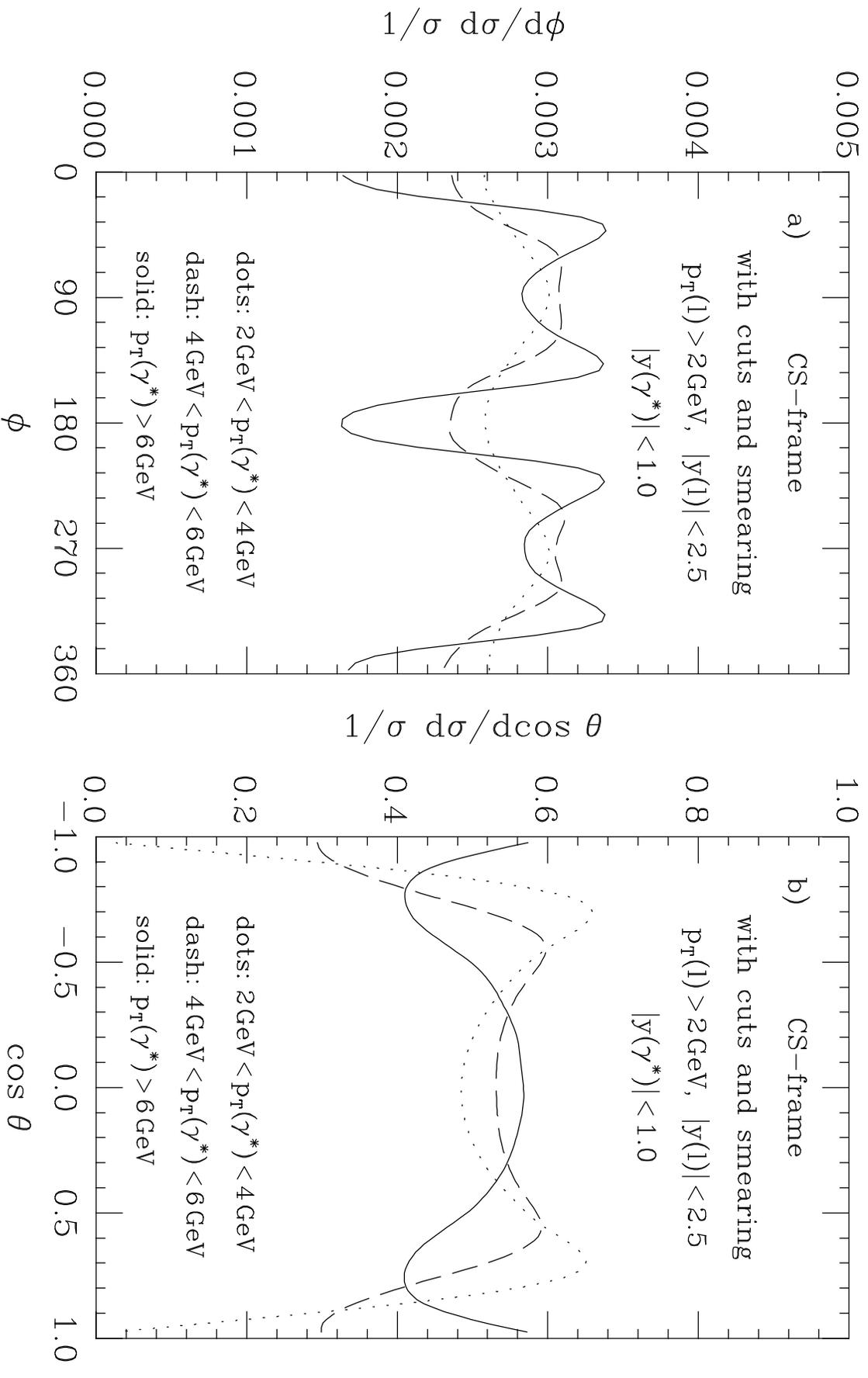
\item[{\bf Fig.~6}]
Same as Fig.~5  but with smearing and the cuts
$p_T^{}(\ell) > 2$~GeV, $|y(\ell)| < 2.5$, and $|y(\gamma^*)|<1.0$.

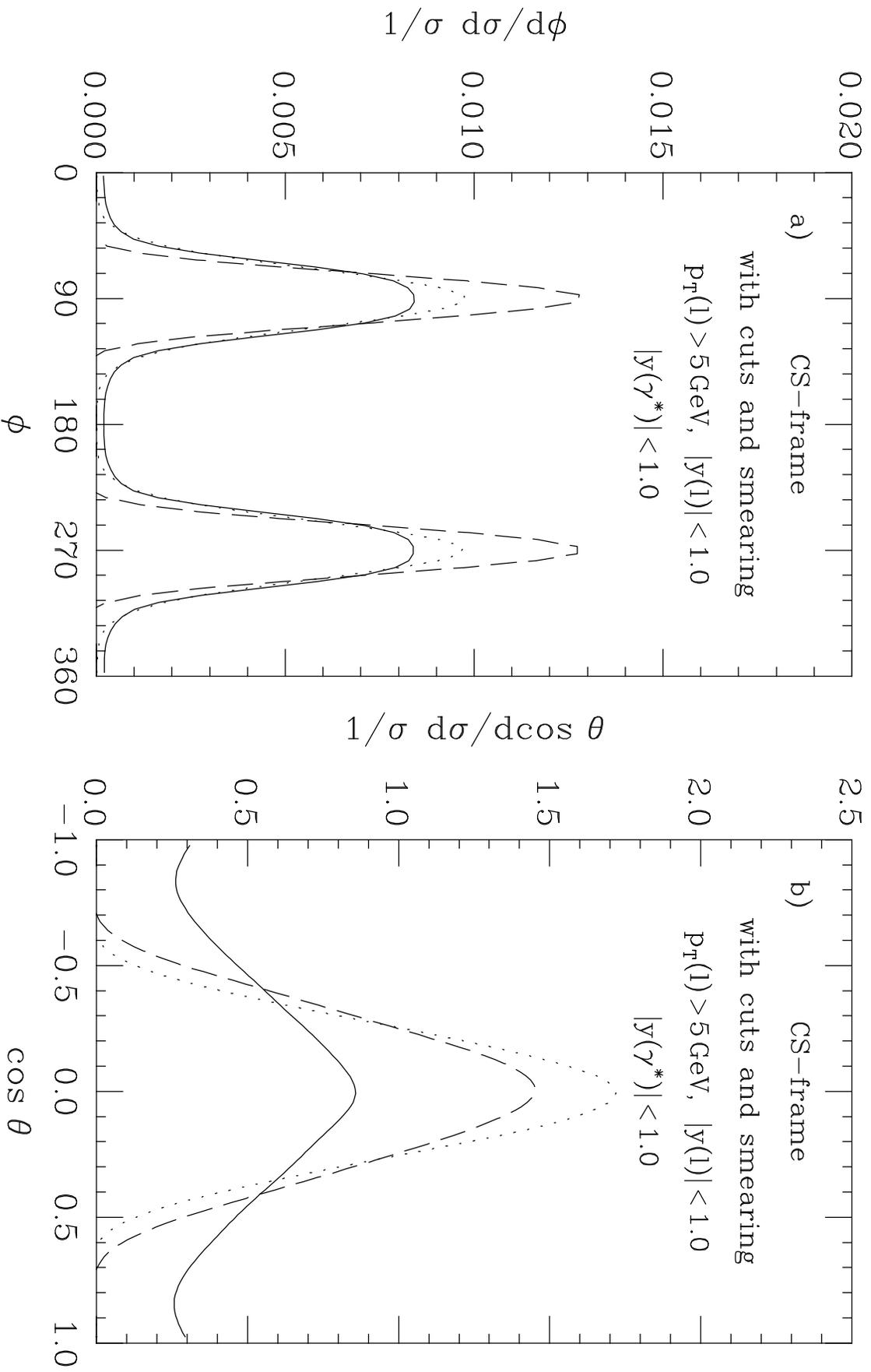
\item[{\bf Fig.~7}]
Same as Fig.~5  but with smearing and the cuts
$p_T^{}(\ell) > 5$~GeV, $|y(\ell)| < 1.0$, and $|y(\gamma^*)|<1.0$.

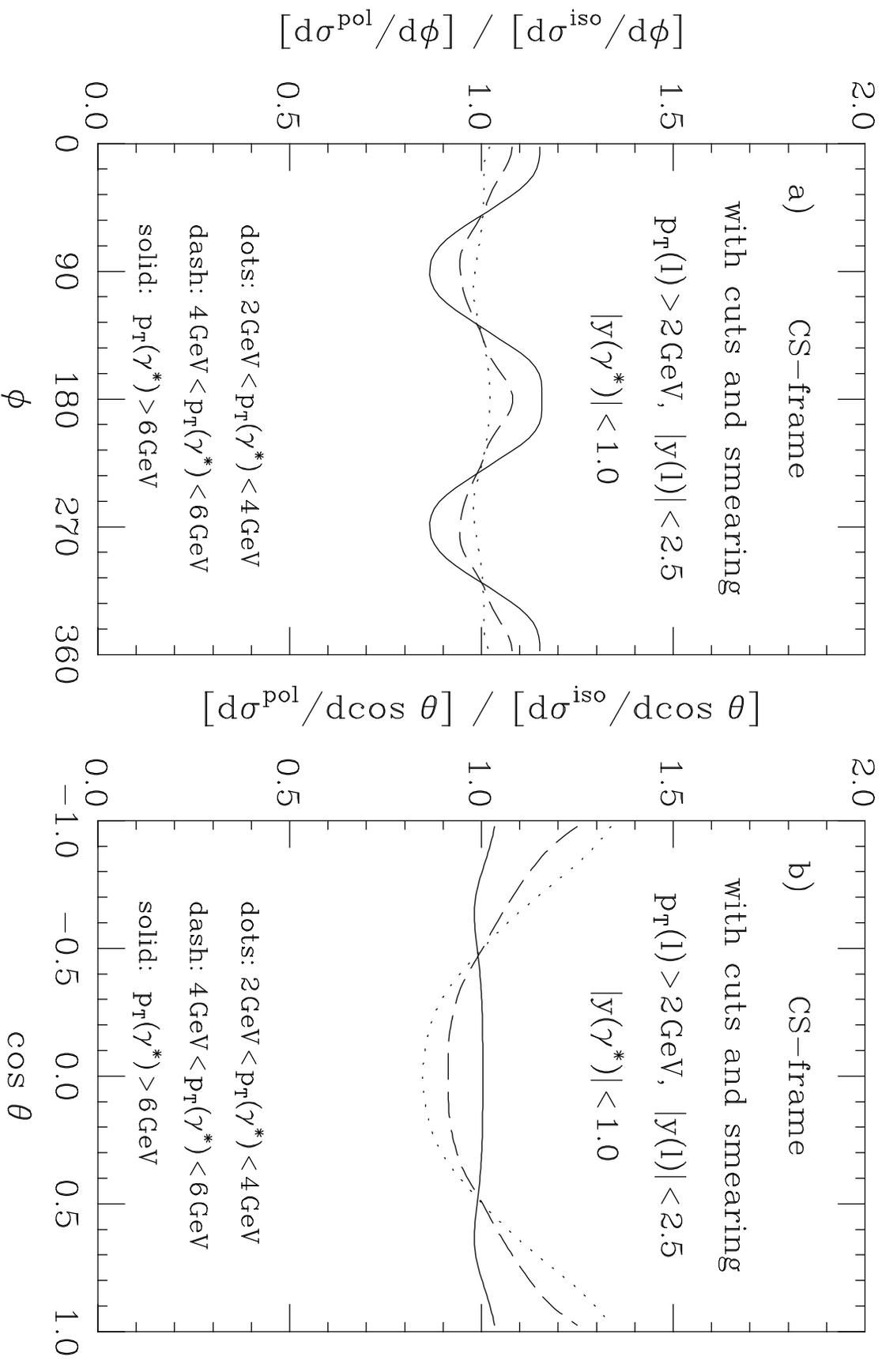
\item[{\bf Fig.~8 }]
Ratios of distributions in the CS frame  obtained with full
polarization to those obtained with isotropic decay of the
$\gamma^*$ for the same  $p_T^{}(\gamma^*)$ bins as in Fig.~5.  Parts a)
and b) are the ratios for the $\phi$ and $\cos\theta$ distributions,
respectively. Energy resolution smearing and the cuts $p_T^{}(\ell) >
2$~GeV, $|y(\ell)| < 2.5$, and
$|y(\gamma^*)|<1.0$ are included.

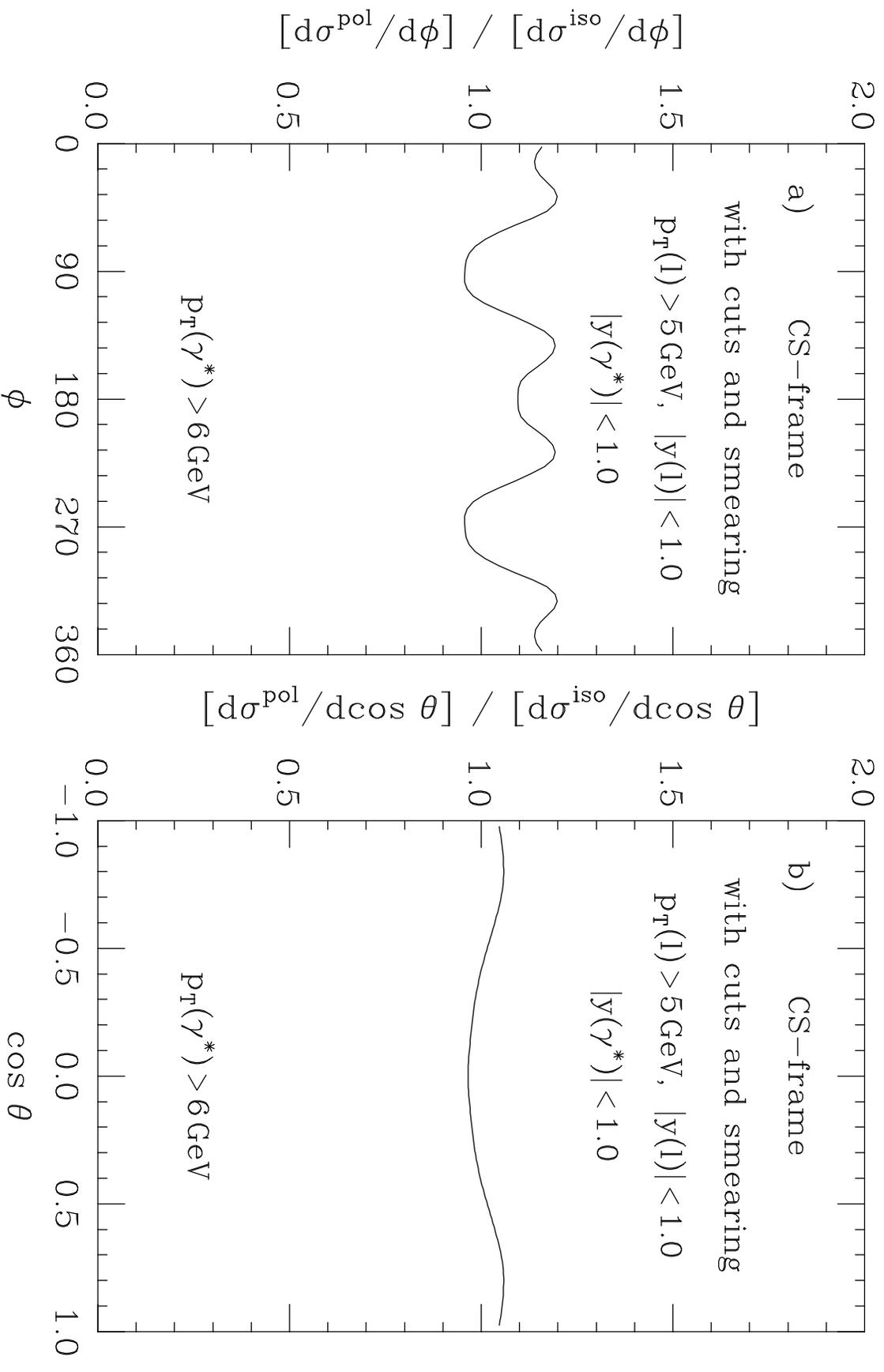
\item[{\bf Fig.~9 }]
Same as Fig.~8 but with the  cuts
$p_T^{}(\ell) > 5$~GeV, $|y(\ell)| < 1.0$,
and $|y(\gamma^*)|<1.0$.
Only the bin for $p_T^{}(\gamma^*) > 6$~GeV is shown.

\end{itemize}

\newpage
\appendix{NLO Matrix elements}
In this appendix, we present the remaining NLO matrix elements which
do not contribute to $W$ boson  production and are thus not listed in
Ref.~\cite{npb}. It is convenient to calculate covariant projection
cross sections $d\hat{\sigma}^{\beta}\,\,\,
[\beta\in\{U+L,L_1,L_2,L_{12}\}]$ from which one can deduce the
helicity cross sections $d\hat{\sigma}^{\alpha}\,\,\,
[\alpha\in\{U+L,L,T,I\}]$ in
Eqs.~(\ref{borntilde})--(\ref{baumtilde}) for any given $\gamma^*$
rest frame by a transformation matrix $(M)_{\alpha\beta}$. For the CS
frame one has \cite{npb}:
\begin{equation}
\left(
\begin{array}{c}
d\hat{\sigma}^{U+L}\\[3mm]
d\hat{\sigma}^L\\[3mm]
d\hat{\sigma}^T\\[3mm]
d\hat{\sigma}^I\\[3mm]
\end{array}
\right)_{CS}
=
\left(
    \begin{array}{cccc}
1   &  0  & 0 &  0 \\[2mm]
0   &  \frac{1}{4\,\cos^{2}\gamma_{CS}}
         &  \frac{1}{4\,\cos^{2}\gamma_{CS}}
         &  \frac{-1}{4\,\cos^{2}\gamma_{CS}}
          \\[3mm]
\frac{1}{2}
         &   -\frac{(1+\cos^{2}\gamma_{CS})}
                 {8\,\sin^{2}\gamma_{CS}\cos^{2}\gamma_{CS}}
         &   -\frac{(1+\cos^{2}\gamma_{CS})}
                 {8\,\sin^{2}\gamma_{CS}\cos^{2}\gamma_{CS}}
        &   \frac{(1-3\cos^{2}\gamma_{CS})}
            {8\,\sin^{2}\gamma_{CS}\cos^{2}\gamma_{CS}}
                        \\[3mm]
0        & \frac{1}{4\sqrt{2}\,\sin\gamma_{CS}\cos\gamma_{CS}}
        & \frac{-1}{4\sqrt{2}\,\sin\gamma_{CS}\cos\gamma_{CS}}
        & 0   \\
    \end{array}
\right)
\
\left(
\begin{array}{c}
d\hat{\sigma}^{U+L}\\[3mm]
d\hat{\sigma}^{L_1}\\[3mm]
d\hat{\sigma}^{L_2}\\[3mm]
d\hat{\sigma}^{L_{12}}\\[3mm]
\end{array}
\right)
\label{csmatrix}
\end{equation}
where
\begin{equation}
\cos\gamma_{CS}=
\sqrt{\frac{Q^2 s}{(t-Q^2)(u-Q^2)}}\>,
\hspace{2cm}
\sin\gamma_{CS}=-\sqrt{1-\cos^2\gamma_{CS}}\>.
\end{equation}

The results for the  GJ frame  can be obtained from:
\begin{equation}
\left(
\begin{array}{c}
d\hat{\sigma}^{U+L}\\[3mm]
d\hat{\sigma}^L\\[3mm]
d\hat{\sigma}^T\\[3mm]
d\hat{\sigma}^I\\[3mm]
\end{array}
\right)_{GJ}
=
\left(
    \begin{array}{cccc}
1   &  0  & 0 &  0 \\[2mm]
0   &  1  & 0 &  0 \\[3mm]
\frac{1}{2}
         &   -\frac{(1+\cos^{2}\gamma_{GJ})}
                 {2\sin^2\gamma_{GJ}}
         &   -\frac{1}{\sin^2\gamma_{GJ}}
        &   \frac{\cos\gamma_{GJ}}
            {\sin^2\gamma_{GJ}}
                        \\[3mm]
0        & \frac{-\cos\gamma_{GJ}}{\sqrt{2}\,\sin\gamma_{GJ}}
        & 0
        &  \frac{1}{2\sqrt{2}\,\sin\gamma_{GJ}}   \\
    \end{array}
\right)
\
\left(
\begin{array}{c}
d\hat{\sigma}^{U+L}\\[3mm]
d\hat{\sigma}^{L_1}\\[3mm]
d\hat{\sigma}^{L_2}\\[3mm]
d\hat{\sigma}^{L_{12}}\\[3mm]
\end{array}
\right)
\label{gjmatrix}
\end{equation}
and
\begin{equation}
\cos\gamma_{GJ}=
 {1-\frac{2Q^2 s}{(t-Q^2)(u-Q^2)}}\>,
\hspace{2cm}
\sin\gamma_{GJ}=-\sqrt{1-\cos^2\gamma_{GJ}}\>.
\end{equation}
All of the partonic projection matrix elements $T_{ab}^{\beta}$ are
listed in Ref.~\cite{npb}, however, there are some interference
contributions which do not contribute to $W$ boson production because
of charge conservation. The analytical results for these additional
partonic projection matrix elements are listed here.\\[5mm]
{I) Diagrams $2(F_{5}+F_{6})^{\ast}(F_{7}+F_{8})$
for $q\bar{q} \rightarrow \gamma^{*}q\bar{q}$}\\[2mm]
The diagrams are shown in Fig.~8 of Ref.~\cite{npb}. The result for
the interference of these diagrams differs for vector-vector (relevant
for $\gamma^{\ast}$ production) and axial-axial  (relevant for $Z$
boson production) couplings. They do not contribute to $W$ boson
production. The vector-vector coupling contribution for the projection
matrix elements $T_{q\bar{q}}^{\beta}$ for these diagrams are denoted
by $D_{cd\,VV}^{\beta}$. All quantities in the following formulae are
defined in appendix~F  of Ref.~\cite{npb}.
\begin{eqnarray*}
D_{cd\,VV}^{U+L}(s,t,u,Q^2)&=&
   \left(\frac{1}{2}\right)
 \left\langle\mbox{\rule[0cm]{0cm}{6mm}}\right.
    \frac{s(s_2^{2}-ut)}{ut}d_{t}d_{u} - \frac{u+t}{2ut}
    - (s+s_2-Q^{2})\frac{(u-t)^{2}}{2ut\lambda^{2}}
            \\[2mm]
&& \hspace{-3cm}
+ \,\frac{f_{\lambda}}{\lambda}
    \left[  (3s+2s_2)\frac{u+t}{2ut} + 2({s-Q^{2}})d_{s}
    - s(u+t)\frac{(u-t)^{2}}{2ut\lambda^{2}}  \right]
            \\[2mm]
&& \hspace{-3cm}
-\,f_{tu} \,Q^2 \frac{d_{u}d_{t}}{ut}\left( 2(s_2-t)^{2}
    +s_2^{2}+s^{2}\right)
            \\[2mm]
&& \hspace{-3cm}
-\, f_{\lambda t}\frac{d_{s} d_{u}}{t}\left[
 (s_2-Q^{2})^{2}+(s_2-t)^{2}+s_2^{2}+s^{2}\right]
            \\[2mm]
&& \hspace{-3cm}
+\, f_{st}\frac{d_{st}d_{s}}{ut}
(2Q^{2}-t)\left(t^{2}+2(s_2^{2}+s^{2}-s_2t)\right)
\left.\mbox{\rule[0cm]{0cm}{5mm}}\right\rangle
\,\,+\,
\left(\frac{1}{2}\right)
\left\langle\mbox{\rule[0cm]{0cm}{5mm}}u\leftrightarrow  t\right\rangle
\end{eqnarray*}
\vspace{5mm}
\begin{eqnarray*}
D_{cd\,VV}^{L_{1}}(s,t,u,Q^2)&=&
                \left(\frac{1}{2}\right)
                \left\langle\mbox{\rule[0cm]{0cm}{6mm}}\right.
               \frac{d_{su}}{8u}\left\{\mbox{\rule[0cm]{0cm}{5mm}}\right.
                 H_1^{(1,0)}\,\, u d_{st}d_s
                   \left[ -48s^{4}+
                     8s^{3}(13s_2-7t-4u)\right.
            \\[2mm]
&& \hspace{-3.5cm}
    -\,4s^{2}\left(3s_2(8s_2-9t)-13u(s_2-t)-2u^{2}+5t^{2}\right)
            \\[2mm]
&& \hspace{-3.5cm}
                    -\,2s\left( 2u(20s_2^{2}-27s_2t+8t^{2})
                      -(18s_2^{2}-21s_2t+4t^{2})(2s_2-t)
                      -(9s_2-8t)u^{2}  \right)
            \\[2mm]
&& \hspace{-3.5cm}    \left.\mbox{\rule[0cm]{0cm}{4mm}}
                    -\,(5u-8s_2+3t)(s_2-u)(s_2-t)(u-4s_2+3t) \right]
            \\[2mm]
&& \hspace{-3.5cm}
                  +\,H_1^{(1,0)}\,\, \left[ 4s^{2}(u+3t-2s_2)
                        +2s\left(7s_2(u-t)+3(t^{2}-u^{2})-2(s_2^{2}-ut) \right.
                        \right.
            \\[2mm]
&& \hspace{-3.5cm}   \left.
                        +\,(2s_2-u-t)\left( 7(s_2-u)^{2}-(s_2-t)^{2} \right)
                        \right]
            \\[2mm]
&& \hspace{-3.5cm}
                 + \,H_1^{(1,1)}\,\,(s_2-u)\left[   2 \left(s(u-2s_2)
                                +2(2s_2-u-t)(s_2-u) \right)\right.
            \\[2mm]
&& \hspace{-3.5cm}      \left.
                        +\,d_{st}u \left(4s^{2}-2s(u+t-s_2)
                            +(s_2-u)(u+s_2-2t)\right)  \right]
            \\[2mm]
&& \hspace{-3.5cm}
                 +\,H_1^{(1,2)}\,\,(s_2-u)^{2}
                      \left[ (2s_2-u-t)
                        + ud_s d_{st}\left( -4s^{2}-2s(3u-4s_2+t)
                        \right.\right.
            \\[2mm]
&& \hspace{-3.5cm}    \left.\left.
            +\,(s_2-t)(3u-4s_2+t) \right)  \right]
                 + H_1^{(1,3)}\,\, u d_{st}(s_2-u)^{3}
            \\[2mm]
&& \hspace{-3.5cm}
                 +\,H_2^{(1,1)}\,\, 2(s_2-t)
                   \left[(4s_2-2u-3t)s -(s_2-t)(2s_2-u-t)
                   \right.
            \\[2mm]
&& \hspace{-3.5cm} \left.
                   +\, \frac{u d_{st}}{2}\left( 16s^{2}+2s(3u-11s_2+7t)
                      +2u(u-7s_2)+t(10u+t-12s_2)+13s_2^{2}\right) \right]
            \\[2mm]
&& \hspace{-3.5cm}
                 +\,H_2^{(1,2)}\,\,(s_2-t)^{2}
                    \left[   -(2s_2-u-t)
                       +ud_{st}d_s\left( -12s^{2}-2s(5u-12s_2+7t)
                       \right.\right.
            \\[2mm]
&& \hspace{-3.5cm}    \left.\left.
                       -\,(3u-4s_2+t)(u-3s_2+2t) \right) \right]
                 + H_2^{(1,3)}\,\, u(s_2-t)^{3}d_{st}
            \\[2mm]
&& \hspace{-3.5cm}
                 + \,
                  2 d_t\left[4s^{2}s_2+2s\left(s_2(2u-3t-s_2)+2t^{2}\right)
                  \right.
            \\[2mm]
&& \hspace{-3.5cm}\left.
                  -\,(s_2-t)\left( (t-s_2)^{2}-3(u-s_2)^{2} \right)\right]
                   + d_{st}\frac{2u}{3}\left( 24s^{2}-4s(4s_2-3t)\right.
            \\[2mm]
&& \hspace{-3.5cm} \left.
                   +\,(u-t)^{2}+16(s_2-u)(s_2-t)\right)
           \left.\mbox{\rule[0cm]{0cm}{5mm}}           \right\}
            \\[2mm]
&& \hspace{-3.5cm}
                       -\, \frac{1}{2u}
                       \left\{\mbox{\rule[0cm]{0cm}{5mm}}\right.
                 f_{tu}2d_{su}d_{st}\left( (s_2-u)(s_2-t)-ss_2 \right)
                       (s-u+s_2)
            \\[2mm]
&& \hspace{-3.5cm}
                +\, f_{su}us(s-u+s_2)d_sd_{st}
                - f_{\lambda u}d_{su}(s-u+s_2)(s_2-u+d_{s}us)
            \\[2mm]
&& \hspace{-3.5cm}
                - \,f_{st}d_{su}s\left(u-s-s_2+d_{s}u(s_2-u-3s)\right)
            \\[2mm]
&& \hspace{-3.5cm}
  + \,f_{\lambda t}d_sd_{st}u\left((Q^2-s_{2})(s_2-t)-s^{2}\right)
         \left.\mbox{\rule[0cm]{0cm}{5mm}}\right\}
         \left.\mbox{\rule[0cm]{0cm}{6mm}}\right\rangle
\end{eqnarray*}
\vspace{3mm}
\begin{eqnarray*}
D_{cd\,VV}^{L_{2}}(s,t,u,Q^2)&=&D_{cd\,VV}^{L_{1}}(s,u,t,Q^2)
\end{eqnarray*}
\vspace{3mm}
\begin{eqnarray*}
D_{cd\,VV}^{L_{12}}(s,t,u,Q^2)&=&
\left(\frac{1}{2}\right)
 \left\langle\mbox{\rule[0cm]{0cm}{5mm}}\right.
        \frac{-d_{su}d_{st}}{8ut}   \left\{\mbox{\rule[0cm]{0cm}{5mm}}\right.
       - H_1^{(1,0)}\,\, d_s\left[
         8s^{4}(3u^{2}-5ut-6us_2)\mbox{\rule[0cm]{0cm}{4mm}}\right.
            \\[2mm]
&& \hspace{-3.5cm}
      +\, 4s^{3} \left( 4u^{2}(3u-2t)-13us_2(3u-t)+30us_2^{2} \right)
            \\[2mm]
&& \hspace{-3.5cm}
      +\, 2s^{2}\left( 2u^{3}(8u+9t)-u^{2}(73us_2+22t^{2})
                + 5u^{2}s_2(9t+20s_2)-12us_2^{2}(4t+3s_2)  \right)
            \\[2mm]
&& \hspace{-3.5cm}
      +\,  s \left( 3u^{4}(3u+t)-4u^{3}(13us_2+19t^{2})\right.
            \\[2mm]
&& \hspace{-3.5cm} \left.
           +\,12u^{2}ts_2(17t+8u)+8u^{2}s_2^{2}(11u-52t)
           -4us_2^{3}(7u-49t+6s_2)  \right)
            \\[2mm]
&& \hspace{-3.5cm}
      +\,  u^{5}(u+2t-7s_2)-3u^{4}(9t^{2}-4s_2^{2})
        +5u^{3}ts_2(5u+46t)+10u^{2}s_2^{3}(u-4s_2)
            \\[2mm]
&& \hspace{-3.5cm} \left.
        +\,8us_2^{4}(3s_2-14t)
        +2u^{2}ts_2^{2}(179s_2-129t-93u)-32u^{3}t^{3}
      \mbox{\rule[0cm]{0cm}{4mm}}          \right]
            \\[2mm]
&& \hspace{-3.5cm}
     +\,  H_1^{(1,1)}\,\,
        2(s_2-u)\left[\mbox{\rule[0cm]{0cm}{4mm}}\right.
         3s^{2}u(u-3t-2s_2)
        +  s\left( 4u^{2}(u-3s_2)\right.
            \\[2mm]
&& \hspace{-3.5cm} \left.
        -\,3ut(u+t-4s_2)+4s_2^{2}(2u-t)
        +2t^{2}s_2 \right)
        +  u^{3}(u+t-4s_2)
            \\[2mm]
&& \hspace{-3.5cm}
        +\,2ut(2s_2-u)(4t-3s_2)
        -3t^{2}(ut+2s_2^{2})
           +us_2^{2}(5u-2s_2)+2ts_2(t^{2}+2s_2^{2})
           \left.\mbox{\rule[0cm]{0cm}{4mm}} \right]
            \\[2mm]
&& \hspace{-3.5cm}
    -\, H_1^{(1,2)}\,\, d_s(s_2-u)^{2}\left[\mbox{\rule[0cm]{0cm}{4mm}}\right.
          2s^{2}\left( u(u-8t-2s_2)-t(t-2s_2) \right)
            \\[2mm]
&& \hspace{-3.5cm}
       +\,  s  \left( 3(u^{3}-t^{3})+10s_2(t^{2}-u^{2})
              +8s_2(4ut+us_2-ts_2)-ut(17u+15t) \right)
            \\[2mm]
&& \hspace{-3.5cm}
       +\,  5s_2(t^{3}-u^{3})+u^{2}(u^{2}+9ts_2+8s_2^{2})
          -t^{2}(t^{2}+6ut-23us_2+10u^{2})
            \\[2mm]
&& \hspace{-3.5cm}
        -\,4us_2^{2}(s_2+4t)+4ts_2^{2}(s_2-2t)
        \left.\mbox{\rule[0cm]{0cm}{4mm}}\right]
            \\[2mm]
&& \hspace{-3.5cm}
      -\, H_1^{(1,3)}\,\, 2ut(s_2-u)^{3}
          \,- 4s^{2}\left[u(3t-2u+s_2)+2s_2^{2}-ss_2\right]
            \\[2mm]
&& \hspace{-3.5cm}
       +\,  \frac{2}{3}s\left[ 3u^{2}(t+5u)-6s_2(6t^{2}-s_2^{2})+22s_2ut\right.
       - \frac{2}{3} u^{2}t(17t-47s_2)
            \\[2mm]
&& \hspace{-3.5cm}
       -\,2u^{3}(3s_2-u)
       -\frac{4}{3}u\left(14ts_2^{2}
          -3s_2^{3}+tu^{2}\right)
     \left.\mbox{\rule[0cm]{0cm}{6mm}}\right\}
            \\[2mm]
&& \hspace{-3.5cm}
            +\, \frac{d_{st}}{2ut}
     \left\{\mbox{\rule[0cm]{0cm}{6mm}}\right.
      f_{tu}d_{su}\left[ s(Q^2-s_2-2s)+s_2(Q^2-s)-2ut \right]
                         \left[(s_2-u)(s_2-t)-ss_2\right]
            \\[2mm]
&& \hspace{-3.5cm}
         +\,  f_{su}sd_s\left[
              2s^{2}(u-2s_2)
            +  s\left(u(u+5t)-2s_2(u+2t)\right)\right.
            \\[2mm]
&& \hspace{-3.5cm} \left.
           -\, 4s_2^{2}(Q^2-s)-ut(2u-t-3s_2)+u^{2}s_2   \right]
            \\[2mm]
&& \hspace{-3.5cm}
          +\, f_{\lambda t}d_{su}\left[
            ss_2(2s_2-u-t)+stu + (s_2-t)(2t-s_2)(t+u-2s_2) \right]
                    \left.\mbox{\rule[0cm]{0cm}{6mm}}\right\}
                    \left.\mbox{\rule[0cm]{0cm}{6mm}}\right\rangle
            \\[2mm]
&& \hspace{-3.5cm}
       +\,
      \left(\frac{1}{2}\right)
   \left\langle\mbox{\rule[0cm]{0cm}{5mm}}  u\leftrightarrow  t\right\rangle
\end{eqnarray*}
{II) Diagrams $2(H_{1}+H_{2})^{\ast}(H_{3}+H_{4})$
                    and $2(H_{5}+H_{6})^{\ast}(H_{7}+H_{8})$
for $qq \rightarrow \gamma^{*}qq$ }\\[2mm]
The diagrams are shown in Fig.~9 of Ref.~\cite{npb}. The vector-vector
coupling contribution for the projection  matrix elements
$T_{qq}^{\beta}$ for the diagrams
$2(H_{1}+H_{2})^{\ast}(H_{3}+H_{4})$     and
$2(H_{5}+H_{6})^{\ast}(H_{7}+H_{8})$ are denoted by
$E_{ab\,VV}^{\beta}$ and $E_{cd\,VV}^{\beta}$, respectively, and can
be obtained from the matrix elements $D_{cd\,VV}^{\beta}$ listed
above:
\begin{eqnarray*}
E_{ab\,VV}^{\beta}(s,t,u,Q^2)&=&E_{cd\,VV}^{\beta}(s,u,t,Q^2)
                           \,=\, -D_{cd\,VV}^{\beta}(s,u,t,Q^2) \>.
\end{eqnarray*}

\def\npb#1#2#3{{\it Nucl. Phys. }{\bf B #1} (#2) #3}
\def\plb#1#2#3{{\it Phys. Lett. }{\bf B #1} (#2) #3}
\def\prd#1#2#3{{\it Phys. Rev. }{\bf D #1} (#2) #3}
\def\prl#1#2#3{{\it Phys. Rev. Lett. }{\bf #1} (#2) #3}
\def\prc#1#2#3{{\it Phys. Reports }{\bf C #1} (#2) #3}
\def\pr#1#2#3{{\it Phys. Reports }{\bf #1} (#2) #3}
\def\zpc#1#2#3{{\it Z. Phys. }{\bf C #1} (#2) #3}
\def\ptp#1#2#3{{\it Prog.~Theor.~Phys.~}{\bf #1} (#2) #3}
\def\nca#1#2#3{{\it Nouvo~Cim.~}{\bf A #1} (#2) #3}
\newpage
\sloppy
\raggedright

\end{document}